\documentclass[aps,twocolumn,floatfix,superscriptaddress]{revtex4-1}
\usepackage[british]{babel}
\usepackage[utf8]{inputenc}
\usepackage{lmodern}
\usepackage[T1]{fontenc}
\usepackage{hyperref}

\usepackage{mathtools} 
\usepackage{amsfonts} 

\usepackage{tikz}
	\usetikzlibrary{decorations.markings} 

\usepackage{float}
\usepackage{placeins}

\usepackage{booktabs} 
\usepackage{tabularx} 
\usepackage{multirow} 

\DeclareMathOperator{\Up}{
	\,
	\tikz[baseline={([yshift=-.5*10pt*0.6]current bounding box.center)},scale=0.6,decoration={markings, mark=at position 0.7 with {\arrow[scale=1.25,>=stealth]{>}}}]{ 
		\draw[postaction=decorate] (0,0) -- (0,.6); }
	\,
}
\DeclareMathOperator{\Down}{
	\,
	\tikz[baseline={([yshift=-.5*10pt*0.6]current bounding box.center)},scale=0.6,decoration={markings, mark=at position 0.7 with {\arrow[scale=1.25,>=stealth]{>}}}]{ 
		\draw[postaction=decorate] (0,.6) -- (0,0); }
	\,
}

\begin{document}

\title{A numerical study of the \textit{F}-model with domain-wall boundaries}

\author{Rick Keesman}\affiliation{Instituut-Lorentz, Universiteit Leiden, Niels Bohrweg 2, 2333 CA Leiden, The Netherlands}

\author{Jules Lamers}\affiliation{Department of Mathematical Sciences, Chalmers University of Technology and University of Gothenburg, SE-412 96 Göteborg, Sweden}\affiliation{\textup{Corresponding author (\texttt{julesl@chalmers.se})}}

\begin{abstract}
	\noindent We perform a numerical study of the \textit{F}-model with domain-wall boundary conditions. Various exact results are known for this particular case of the six-vertex model, including closed expressions for the partition function for any system size as well as its asymptotics and leading finite-size corrections. To complement this picture we use a full lattice multi-cluster algorithm to study equilibrium properties of this model for systems of moderate size, up to $L=512$. We compare the energy to its exactly known large-$L$ asymptotics. We investigate the model's infinite-order phase transition by means of finite-size scaling for an observable derived from the staggered polarization in order to test the method put forward in our recent joint work with Duine and Barkema. In addition we analyse local properties of the model. Our data are perfectly consistent with analytical expressions for the arctic curves. We investigate the structure inside the temperate region of the lattice, confirming the oscillations in vertex densities that were first observed by Sylju{\aa}sen and Zvonarev, and recently studied by Lyberg \textit{et al}. We point out `(anti)ferroelectric' oscillations close to the corresponding frozen regions as well as `higher-order' oscillations forming an intricate pattern with saddle-point-like features.
\end{abstract}

\maketitle

\section{Introduction}\label{sec1}

The \textit{F}-model for antiferroelectric materials~\cite{Rys_63} is a special case of the six-vertex, or ice-type, model that exhibits an infinite-order phase transition (IOPT)~\cite{Lie_67b}. Amongst others, studying the \textit{F}-model may thus be instructive to get a better grasp of the well-known IOPT of the two-dimensional \textit{XY}-model as it offers a more simple setting in which the microscopic degrees of freedom are discrete. By definition, at an IOPT the physics of a system does not change as abruptly as it does for finite-order phase transitions, which makes numerical investigations a rather subtle issue. In \cite{KLDB_16}, together with Duine and Barkema, we proposed a new observable for numerical studies of IOPTs: the logarithmic derivative of the (smooth but not analytic) order parameter for the IOPT. By construction this quantity exhibits a peak at the critical --- or rather `transition' --- temperature~$\beta_\text{c}$ of the model, which makes it a suitable candidate for the analysis of the physics near the IOPT. We used a finite-size scaling analysis to compare the performance of our observable with that of other observables commonly used in the literature, focussing on the \textit{F}-model with periodic boundary conditions (PBCs) in both directions. In the present work we test the observable in a different, yet closely related, setting. At the same time this allows us to investigate other intriguing features of the \textit{F}-model, such as the dependence of its thermodynamics, i.e.~the behaviour at asymptotically large system size, on the boundary conditions.

The microscopic degrees of freedom of the six-vertex model are arrows pointing in either direction along the edges of a square lattice. Around each vertex the arrows have to obey the so-called ice rule, which turns out to be rather restrictive~\footnote{To see that the ice rule is crucial here consider the eight-vertex model, where the ice rule is slightly relaxed. This model cannot be tackled with a straightforward Bethe-ansatz analysis, and its thermodynamics are insensitive to the choice of boundary conditions, cf.~\cite{BKW_73} below.}. On the one hand this condition famously allows for a Bethe-ansatz analysis that provides \emph{exact} results, see e.g.~\cite{Lam_14} and references therein, in the thermodynamic limit. On the other hand it causes the model's thermodynamics to depend on the \emph{choice of boundary conditions} used at the intermediate analysis for finite size~\cite{BKW_73,KZ_00,Zin_02u}. (In fact, this phenomenon in the context of graphene~\cite{JJB_16} originally motivated \cite{KLDB_16} and the present work.) PBCs are commonly employed and are compatible with the translational invariance that is present for infinite systems. For the six-vertex model this choice is important for the Bethe ansatz, cf.~\cite{Lie_67b}. This choice was also used in our previous work~\cite{KLDB_16}. The same thermodynamic behaviour is obtained for `free' and (conjecturally) `N\'eel' boundary conditions, where the arrows on the external edges are respectively left free or fixed to alternate~\cite{BKW_73,TRK_15b}. This is not true for `ferroelectric' boundary conditions, where the arrows at the boundary all point e.g.~up or to the right, but with a single allowed microstate the resulting thermodynamics is trivial.

An interesting intermediate case is provided by \emph{domain-wall boundary conditions} (DWBCs), where on two opposite boundaries the arrows all point outwards whereas on the other two boundaries all arrows point inwards. Such boundary conditions first appeared in the calculation of norms of Bethe vectors in the quantum inverse-scattering method (QISM) in the work of Korepin~\cite{Kor_82}. Indeed, the QISM allows for an algebraic construction of the Bethe-ansatz vectors for the Heisenberg \textsc{xxx} and \textsc{xxz} spin chains and the six-vertex model with PBCs. These algebraic Bethe-ansatz vectors simultaneously diagonalize the spin-chain Hamiltonian and the transfer matrix of the six-vertex model provided the parameters featuring in the ansatz obey constraints known as the Bethe-ansatz equations, see e.g.~\cite{Lam_14}. The partition function of the six-vertex model with DWBCs, also known as the \emph{domain-wall partition function}, is related to the norm of the algebraic Bethe-ansatz vectors~\cite{Kor_82}. Later this quantity was found to have applications ranging from the combinatorics of alternating-sign matrices~\cite{MRR_83,Kup_95} (see also the book~\cite{Bre_99}) to one-dimensional quantum systems with inhomogeneous initial conditions that are relevant for cold-atom physics~\cite{AD+_16} to three-point amplitudes in $\mathcal{N}=4$ super Yang--Mills theory~\cite{Kos_12,JK+_16}.

The domain-wall partition function admits a concise closed expression for all system sizes~\cite{Ize_87,ICK_92}. From this the infinite-size asymptotics can be found~\cite{KZ_00,Zin_00}, as well as the form of the leading finite-size corrections~\cite{BF_06,BL_09a,BB_12,BL_13}. The phase diagram of the six-vertex model has the same form for PBCs and DWBCs, but the details are different~\cite{Bax_07,KZ_00,Zin_00}; for example, even though the \textit{F}-model exhibits an IOPT in both cases, the free energy per site of the \textit{F}-model is larger for DWBCs than for PBCs.  In the past decade or so DWBCs have also attracted considerable attention in relation to the arctic-curve phenomenon: they lead to coexisting phases that are spatially separated, with an \emph{arctic curve} separating the `frozen' (ordered) and `temperate' spatial regions. This has been investigated from numerical~\cite{SZ_04,AR_05,CGP_15,LKV_16u} as well as analytic~\cite{JPS_98u,Joh_05,FS_06,CP_10a,CP_10b,CPZ_10,CP_15,AD+_16} viewpoints.

The remainder of this paper is organized as follows. In Sec.~\ref{sec:sec2} we review the \textit{F}-model with DWBCs, its partition function, and the relevant observables; in particular we give a description of the staggered six-vertex model (cf.~\cite{Bax_73b}) in the framework of the QISM. The Monte Carlo cluster algorithm and data processing are discussed in Sec.~\ref{sec:sec3}. The results are treated in Sec.~\ref{sec:sec4}. We fit the exact asymptotic expressions for the energy, giving best estimates for the free parameters in the finite-size corrections, and perform a finite-size scaling analysis to test our observable at the IOPT. Besides these global averaged properties we use our simulations to examine local properties: the coexisting phases, arctic curves, and the structure in the temperate region of the lattice. We conclude with a summary and outlook in Sec.~\ref{sec:sec5}. In App.~\ref{sec:app} we review the global symmetries of the \textit{F}-model and describe how these can be exploited to sample the full phase space. This work is supplemented by an interactive Mathematica notebook~\footnote{See Supplemental Material at [URL will be inserted by publisher] for an interactive Mathematica notebook that provides more detail on the relation between configurations through symmetries and on the chequerboards in the AF oscillations.} to illustrate some features in more detail.

\section{Theory}\label{sec:sec2}

\subsection{The \textit{F}-model and domain walls}

\begin{figure}[!h]
	\centering
	\includegraphics[width=1.0\linewidth,clip=]{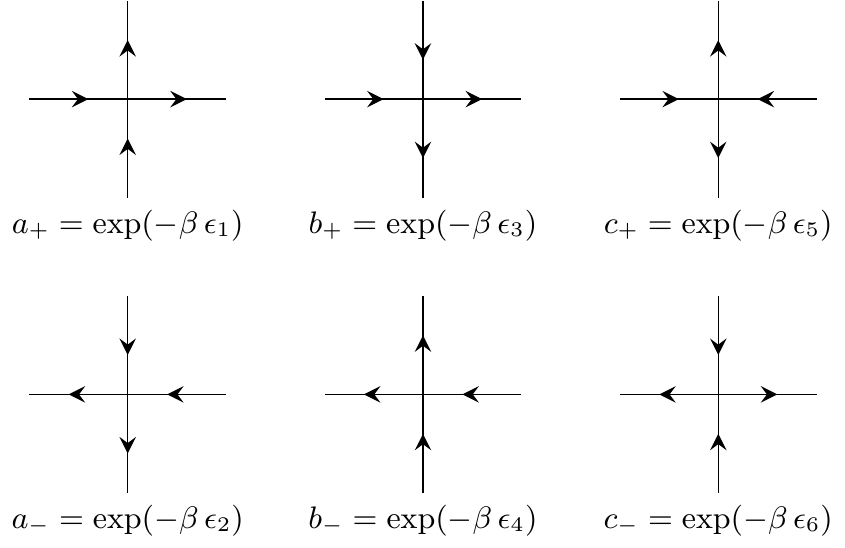}
	\caption{The six vertices allowed by the ice rule and their weights for the six-vertex model. Often one assumes arrow-reversal symmetry: $a_\pm=a$, $b_\pm=b$, $c_\pm=c$. The \textit{F}-model is defined by $a=b<c$.}
	\label{fig:sixvertices}
\end{figure}

The six-vertex model, or (energetic) ice-type model, is a vertex model on a square lattice. The arrows on the edges are restricted by the \emph{ice rule}, which demands that at every vertex two arrows point inwards and two point outwards. This leaves the six allowed vertex configurations shown in Fig.~\ref{fig:sixvertices}. To each such vertex configuration~$i$ one assigns (local) Boltzmann weight $\exp({-\beta}\, \epsilon_i)$, with $\beta:=1/(k_B T)$ the inverse temperature, $k_B>0$ the Boltzmann constant that we put to unity from here on, and $\epsilon_i$ the energy of the vertex configuration. The energy is additive, so the weight of a configuration is the product of these local weights. Summing these over all allowed configurations, subject to some choice of boundary conditions, one obtains the model's partition function.

\begin{figure}[!h]
	\centering
	\includegraphics[width=.55\linewidth,clip=]{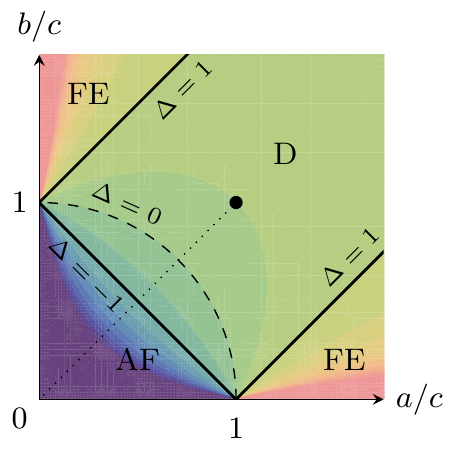}
	
	\caption{The phase diagram of the six-vertex model, parametrized by the ratios $a/c$ and $b/c$ since common rescalings of the vertex weights only yield an overall factor for the partition function. The colours show contours for $\Delta$ at steps of $1/2$ for $-4\leq \Delta\leq 4$. The dashed arc is the so-called free-fermion line. The dotted line corresponds to the \textit{F}-model, with an infinite-order phase transition between the antiferroelectric~(AF) and disordered~(D) phases. The thick dot is the ice point $a=b=c$, which can be interpreted as $\beta=0$. There are two ferroelectric~(FE) phases.}
	\label{fig:fullpd}
\end{figure}

The \textit{F}-model can be obtained by taking  $\epsilon_1=\epsilon_2=\epsilon_3=\epsilon_4$ and $\epsilon_5=\epsilon_6$ such that the corresponding vertex weights are related by $a=b=\exp({-\beta}\,\epsilon)\,c$ for some $\epsilon>0$, making vertices $5$ and~$6$ energetically favourable. Interestingly, this model has experimental realizations using artificial spin ice~\cite{NMS_13}\footnote{We thank P.~Henelius for making us aware of this}. The phase diagram is shown in Fig.~\ref{fig:fullpd}. For low enough temperatures the system is in the antiferroelectric~(AF) phase. As temperature increases there is a transition to the disordered~(D) phase. For PBCs the ground state consists of vertices $5$ and~$6$ alternating in a chequerboard-like manner; this global AF order persists throughout the AF~phase and is destroyed upon entering the D~phase.

The six-vertex model does not have a thermodynamic limit in the usual sense: the  physical properties of macroscopic systems remain sensitive to the choice of boundary conditions. Rather than imposing PBCs we consider an $L \times L$ portion of the lattice with \emph{domain-wall boundary conditions} (DWBCs), where the arrows on external edges are fixed and point out (inwards) on all horizontal (vertical) edges, say. This change in boundary conditions has several interesting consequences that will be reviewed momentarily. Similarly to the case of PBCs (see e.g.~\cite{Bax_07,LW_72} for reviews) one obtains exact results for the DWBC \textit{F}-model by extending it to the six-vertex model with general vertex weights $a$, $b$, and $c$ as in Fig.~\ref{fig:sixvertices}. The `reduced coupling constant' is defined as
\begin{equation}\label{eq:Delta}
\Delta \coloneqq \frac{a^2 + b^2 - c^2}{2\,ab} \ .
\end{equation}

The phase diagram looks again like in Fig.~\ref{fig:fullpd}. At high temperatures the system is in the D~phase, $-1<\Delta<1$. As the temperature is lowered it transitions into the AF~phase, $\Delta<-1$, or one of the two the ferroelectric~(FE) phases, $\Delta>1$, depending on the ratio $a:b:c$. The D--AF phase transition is of infinite order for PBCs~\cite{Lie_67b} as well as DWBCs (cf.~the end of the following subsection)~\cite{Zin_00,BB_12}, while those between the D and FE phases are of first order for PBCs~\cite{Lie_67c} but of second order for DWBCs~\cite{KZ_00,BL_09b}.

In the FE~phase the DWBCs are compatible with the FE~order, while for $\Delta<1$ (including the \textit{F}-model) the boundaries raise the free energy per site with respect to the case of PBCs. Zinn-Justin~\cite{Zin_02u} suggested that this can be understood as a consequence of coexisting phases that are spatially separated. This phenomenon had also been found for various choices of fixed boundary conditions for the ice model ($a=b=c$) before~\cite{Elo_99}. Through the ice rule the DWBCs induce ordered regions that extend deep into the bulk, and translational invariance is lost even far away from the boundary. For example, the ground state is no longer a chequerboard-like configuration of vertices $5$ and~$6$ as for PBCs, which would after all lead to alternating arrows along the boundary. Instead the DWBC ground state consists of a central diamond-shaped area with AF order (see also Fig.~\ref{fig:Cdensity}~(a) below), consisting of vertices 5 and~6 like before, enclosed by corners that each possess FE order, containing a homogeneous configuration of one of the vertices 1 to~4. (When $L$ is even there are two ground-state configurations of this form.) The domain walls thus raise the ground-state energy per site in the thermodynamic limit from $0$ for PBCs to $\epsilon/2$ for DWBCs. When the temperature becomes nonzero a disordered region appears that separates the regions of AF and FE order, and above the critical temperature the region with AF order disappears to leave a central disordered region surrounded by FE-ordered regions~\cite{SZ_04,AR_05}. There are sharp transitions between the regions, and the curves separating the `frozen' (AF or FE ordered) and `temperate' regions in the scaling limit (i.e.~let $L\to\infty$ while decreasing the lattice spacing to keep total system size fixed) are known as \emph{arctic curves}. These curves have four contact points with the boundary, which for the \textit{F}-model lie in the middle of each side~\cite{CP_10b}. For the `free-fermion point'~$\Delta=0$ the arctic curve is a circle~\cite{JPS_98u} up to fluctuations of order $\sim L^{1/3}$ governed by an Airy process \cite{Joh_05,FS_06}. The arctic curve has also been conjectured for $|\Delta|<1$ \cite{CP_10a,CP_10b} and $\Delta<-1$ \cite{CPZ_10}, where the latter focusses on the curve separating the FE and D regions.

Because we are interested in the \textit{F}-model from now on we focus on the D and AF phases. The following (real) parametrization of the vertex weights are often used in these regimes~\footnote{Computations using quantum integrability for finite size are often done over $\mathbb{C}$, for which $a=\sinh(u+\eta)$, $b=\sinh u$, $c=\sinh\eta$ is another convenient parametrization. Setting $u=-\gamma-t$ and $\eta=2\gamma+i\,\pi$ yields the AF parametrization in~\eqref{eq:parametrizations} up to a common sign; including a factor of $i$ in front of $\gamma$ and $t$ gives that for D up to a common factor of~$i$.}:
\begin{equation}\label{eq:parametrizations}
\text{D: } \begin{cases} a =\sin(\gamma-t) \\ b =\sin(\gamma+t) \\ c =\sin 2\gamma \end{cases}\!\!\!\!\!, \qquad \text{AF: } \begin{cases} a =\sinh(\gamma-t) \\ b =\sinh(\gamma+t) \\ c =\sinh 2\gamma \end{cases}\!\!\!\!\!.
\end{equation}
Here $t \in [{-\gamma},\gamma]$ is called the spectral parameter, while $\gamma\geq 0$ is the crossing parameter, which for the D~phase is further restricted to $\gamma<\pi/2$; it is related to \eqref{eq:Delta} via $\Delta={-\cos2\gamma}$ for D and $\Delta={-\cosh2\gamma}$ for AF. The \textit{F}-model then corresponds to $t=0$, with $\Delta=1-e^{2\beta\epsilon}/2$ or $\gamma$ encoding the temperature as
\begin{equation}\label{eq:temperature}
e^{\beta\epsilon} = \frac{c}{a} = \frac{c}{b} = \begin{cases} 2\cos \gamma\ , \quad\ \, \gamma\in[0,\pi/3] &  \text{(D)} \ , \\ 2\cosh \gamma\ ,  \quad \gamma\geq 0  & \text{(AF)} \ . \end{cases}
\end{equation}
The phase transition of the \textit{F}-model occurs at $\beta_\text{c}\,\epsilon =\ln2$ ($\Delta={-1}$, $\gamma=0$). At this point the parametrization~\eqref{eq:parametrizations} vanishes identically, which can be avoided by simultaneously rescaling the weights to set $c$ equal to unity. At the level of the partition function this may be implemented by keeping \eqref{eq:parametrizations} with $t=0$ but considering the `renormalized' partition function $c^{-L^2} Z_L(a,b,c) = Z_L(a/c,b/c,1)$. We will denote this quantity simply by $Z_L$.

\subsection{The domain-wall partition function}\label{sec:DW_partition_function}

In some sense the six-vertex model with DWBCs is a theorist's dream. Unlike for PBCs, for which exact results are only available for asymptotically large systems, the domain-wall partition function~$Z_L$ can be found exactly for \emph{all} system sizes. In brief the computation goes as follows, see e.g.~\cite{Kup_95} for more details. For the $i$th row ($j$th column) of the lattice one introduces a parameter $u_i$ ($v_j$). This allows one to further extend the model to an inhomogeneous version where the weight~\eqref{eq:parametrizations} at position $(i,j)$ features $u_i-v_j$ instead of $t$. Korepin~\cite{Kor_82} showed that $Z_L$, viewed as a function of the~$u_i$, obeys certain properties that determine it uniquely in the inhomogeneous setting; most importantly there is a recursion relation that expresses $Z_L$ with one $u_i$ specialized to a specific value in terms of $Z_{L-1}$. Izergin~\cite{Ize_87,ICK_92} found a remarkably concise expression in the form of a determinant of an $L\times L$ matrix. Since it meets all Korepin's requirements, Izergin's determinant provides a formula for the domain-wall partition function valid for all~$L$. Upon carefully evaluating the homogeneous limit, $u_i - v_j \to t$ for all $i$ and~$j$, this results in a Hankel determinant:
\begin{equation}\label{eq:Z_finite}
Z_L = \frac{(ab/c)^{L^2}}{\prod_{k=0}^{L-1} (k!)^2} \ \det_{L\times L} M \ , \qquad M_{i,j} \coloneqq \partial_t^{i+j-2} \frac{c}{ab} \ ,
\end{equation}
where the definition of $M_{i,j}$ assumes a parametrization of the form~\eqref{eq:parametrizations}. Specializing this quantity to the ice (or `combinatorial') point~$a=b=c$ (so $\Delta=1/2$) one finds that the number of domain-wall configurations for $L=1,2,\dots$ is $1$, $2$, $7$, $42$, $429$, $7436$, $218348$, \dots~\cite{Kup_95}. For the \textit{F}-model the domain-wall partition function factorizes as $Z_{2L} = 2\, X_{2L} X_{2L+1}$, $Z_{2L+1} = X_{2L+1} X_{2L+2}$ for certain polynomials $X_L$~\cite[Thm.~3]{Kup_95}, cf.~\cite[Thm.~4]{Kup_02}.

Using the explicit results found by Korepin--Izergin the bulk free energy was evaluated in the thermodynamic limit by Korepin and Zinn-Justin~\cite{KZ_00} and Zinn-Justin~\cite{Zin_00}. Prior to that only some special cases in the D~phase were known: the free-fermion point ($\Delta=0$, $\gamma=\pi/4$) corresponding to the 2-enumeration of alternating-sign matrices~\cite[Sec.~6]{MRR_83}, and the ice point ($\Delta=1/2$, $\gamma=\pi/3$) as well as the point $\Delta=-1/2$ ($\gamma=\pi/6$) related to the 3-enumeration of alternating-sign matrices~\cite{Kup_95}. Here we recall that the `$c^2$-enumeration of alternating-sign matrices', cf.~e.g.~\cite{Kup_95,Bre_99}, is given by $c^L \, Z_L(1,1,c)$ since the DWBCs imply that $\#\,c_- = L + \#\,c_+$.

A rigorous and more detailed analysis for the D and AF phases and the corresponding transition, which is most relevant for us, was given by Bleher \textit{et al}.~\cite{BF_06,BL_09a,BB_12,BL_13}. The asymptotic expressions for the domain-wall partition function~$Z_L$, together with the first subleading terms in system size, are as follows for the \textit{F}-model. In the disordered regime one has~\cite{BF_06}
\begin{equation}\label{eq:zD}
Z_{\text{D}}^{\text{asym}} = C_\text{D}(\gamma)\, f_{\text{D}}(\gamma)^{L^2} L^{\kappa(\gamma)}\, [1+O(L^{-\alpha})] \ ,
\end{equation}
where $C_\text{D}(\gamma)>0$ and $\alpha>0$ are unknown (cf.~\cite{Note7} below),
while
\begin{equation}\label{eq:parD}
f_{\text{D}}(\gamma) = \frac{\pi\tan\gamma}{4\gamma} \ , \qquad \kappa(\gamma) =\frac{1}{12}-\frac{2 \gamma^2}{3\pi(\pi-2\gamma)} \ .
\end{equation}
For the antiferroelectric regime one finds~\cite{BL_09a}
\begin{equation}\label{eq:zAF}
Z_{\text{AF}}^{\text{asym}} = C_{\text{AF}}(\gamma)\, f_{\text{AF}}(\gamma)^{L^2} \vartheta_4(L \pi/2)\,[1+O(L^{-1})] \ ,
\end{equation}
with $C_\text{AF}(\gamma)>0$ another unknown normalization factor, and the extensive part of the free energy is
\begin{equation}\label{eq:parAF}
f_{\text{AF}}(\gamma) = \frac{\pi\tanh\gamma}{4\gamma} \,\frac{\vartheta_{1}^{\prime}(0)}{ \vartheta_1(\pi/2)} \ ,
\end{equation}
where $\vartheta_1$ and $\vartheta_4$ denote the Jacobi theta functions with temperature-dependent elliptic nome $q\coloneqq \exp(-\pi^2/2 \gamma)$.

From these exact asymptotics of the domain-wall partition function it can be shown that, as for PBCs, the phase transition is of infinite order~\cite{Zin_00,BB_12}. Indeed, when subtracting the regular part, $(\pi/4\gamma)\tanh\gamma$ [differing from \eqref{eq:parD} only in the parametrization used], from the AF free energy~\eqref{eq:parAF} one is left with an expression that is smooth but exhibits an essential singularity as $\gamma\to0^+$.

\subsection{The staggered polarization}\label{sec:P_0}

An order parameter for the D--AF phase transition is defined as follows. For any microstate~$C$ one can compute the spontaneous staggered polarization~$P_0(C)$. This quantity is a measure of the likeness of $C$ to one of the two AF ground states $C^\prime$ of the system with PBCs. At each vertex the local spontaneous staggered polarization can be defined as $\sum_i \sigma_i \sigma_i^\prime/4$, where the sum is taken over the four edges surrounding the vertex, and $\sigma_i=\pm1$ ($\sigma_i'=\pm1$) depending on whether arrows on those edges point outwards or inwards in $C$ ($C'$). Then $P_0(C)$ is the sum over all these local quantities; since the AF ground state is two-fold degenerate its sign depends on the choice of $C^\prime$ to which $C$ is compared. Additionally, for even~$L$ states come in pairs with equal energy but opposite spontaneous staggered polarization. To avoid cancellation of these contributions one defines the \emph{staggered polarization} as the thermal average $P_0:=\langle \, |P_0(C)| \, \rangle$ of the absolute value of $P_0(C)$. Note that the situation is analogous to what happens for the magnetization in the two-dimensional Ising model.

For the system with PBCs Baxter derived the exact large-$L$ asymptotics of $P_0$ for all temperatures~\cite{Bax_73b}. This quantity becomes smoothly nonzero when the system transitions from the D to the AF phase. Let us assume that it continues to be a valid order parameter for the transition of the system with DWBCs. For this case an expression for $P_0$ that is manageable for all system sizes is not known. We still have
\begin{equation}\label{eq:P_0}
P_0 = \frac{d \ln Z_L^+(s)}{d s}  \Big|_{s=0}\ ,
\end{equation}
where $Z_L(s)$ is the partition function of the \textit{F}-model on an $L\times L$ lattice with DWBCs in the presence of an external staggered electric field of strength~$s\geq 0$. The superscript `+' in~\eqref{eq:P_0} indicates that the absolute value of each coefficient is to be taken in order to prevent the aforementioned cancellation. No analogue of \eqref{eq:Z_finite} is known when $s\neq 0$. Nevertheless the framework of the quantum inverse-scattering method (QISM) does allow for the direct computation of $Z_L(s)$, and thus $P_0$, for low system size. Let us indicate how this works; we refer to \cite{Lam_14} and references therein for more about the QISM.

Let us give a description of the \emph{staggered six-vertex model} based on Baxter~\cite{Bax_73b}. We focus on the homogeneous case; inhomogeneities may be incorporated as usual. View the square lattice as being bipartite by dividing its vertices into two sets in a chequerboard-like manner. The vertex weights from Fig.~\ref{fig:sixvertices} are given by $a_\pm=a$, $b_\pm=b$, while $c_\pm$ is equal to $e^{\pm s}c$ on one sublattice (`black' vertices) and to $e^{\mp s}c$ on the other (`white' vertices). These vertex weights can be encoded in the so-called \emph{\textit{R}-matrix}~\footnote{Note that $R(s) = (\delta_{s/2} \otimes \delta_{-s/2}) \, R(0)\, (\delta_{-s/2} \otimes \delta_{s/2})$ for $\delta_s \coloneqq \exp(s \,\sigma^z/2)$ with $\sigma^z$ the third Pauli matrix, and $R(0)$ the \textit{R}-matrix of the ordinary (zero-field) six-vertex model. By the ice rule one can rewrite $R(s) = (I \otimes \delta_{-s}) \, R(0)\, (\delta_s \otimes I) = (\delta_s \otimes I) \, R(0)\, (\delta_{-s} \otimes I)$ with $I$ the $2\times 2$ identity matrix. Direct horizontal and vertical fields would correspond to $(\delta_h \otimes \delta_v) \, R(0)\, (\delta_h \otimes \delta_v)$, which can be used to compute the direct polarization in a similar fashion.}
\begin{equation}\label{eq:R-matrix}
R(s) = \
\tikz[baseline={([yshift=-.5*10pt*0.6]current bounding box.center)},scale=0.6,font=\scriptsize]{
	\draw (0,1) -- (2,1);
	\draw (1,0) -- (1,2);
	\draw[fill=white]  (1,1) node{$s$} circle (.3);
}
\ = \begin{pmatrix}	a & \color{gray!85}{0} & \color{gray!85}{0} & \color{gray!85}{0} \\ \color{gray!85}{0} & b & e^s \, c & \color{gray!85}{0} \\ \color{gray!85}{0} & e^{-s} \, c & b & \color{gray!85}{0} \\ \color{gray!85}{0} & \color{gray!85}{0} & \color{gray!85}{0} & a \end{pmatrix} ,
\end{equation}
defined with respect to the basis $\lvert
\tikz[baseline={([yshift=-.5*10pt*0.6]current bounding box.center)},scale=0.6,decoration={markings, mark=at position 0.7 with {\arrow[scale=1.25,>=stealth]{>}}}]{ 
	\draw[postaction=decorate] (0,.6) -- (.5,.6);
	\draw[postaction=decorate] (.6,0) -- (.6,.5); }
\,\rangle$, $\lvert
\tikz[baseline={([yshift=-.5*10pt*0.6]current bounding box.center)},scale=0.6,decoration={markings, mark=at position 0.7 with {\arrow[scale=1.25,>=stealth]{>}}}]{ 
	\draw[postaction=decorate] (0,.6) -- (.5,.6);
	\draw[postaction=decorate] (.6,.5) -- (.6,0); }
\,\rangle$, $\lvert
\tikz[baseline={([yshift=-.5*10pt*0.6]current bounding box.center)},scale=0.6,decoration={markings, mark=at position 0.7 with {\arrow[scale=1.25,>=stealth]{>}}}]{ 
	\draw[postaction=decorate] (.5,.6) -- (0,.6);
	\draw[postaction=decorate] (.6,0) -- (.6,.5); }
\,\rangle$, $\lvert
\tikz[baseline={([yshift=-.5*10pt*0.6]current bounding box.center)},scale=0.6,decoration={markings, mark=at position 0.7 with {\arrow[scale=1.25,>=stealth]{>}}}]{ 
	\draw[postaction=decorate] (.5,.6) -- (0,.6);
	\draw[postaction=decorate] (.6,.5) -- (.6,0); }
\,\rangle$ for the `incoming' lines and $\langle\,
\tikz[baseline={([yshift=-.5*10pt*0.6]current bounding box.center)},scale=0.6,decoration={markings, mark=at position 0.7 with {\arrow[scale=1.25,>=stealth]{>}}}]{ 
	\draw[postaction=decorate] (.1,0) -- (.6,0);
	\draw[postaction=decorate] (0,.1) -- (0,.6); }
\,\rvert$, $\langle\,
\tikz[baseline={([yshift=-.5*10pt*0.6]current bounding box.center)},scale=0.6,decoration={markings, mark=at position 0.7 with {\arrow[scale=1.25,>=stealth]{>}}}]{ 
	\draw[postaction=decorate] (.1,0) -- (.6,0);
	\draw[postaction=decorate] (0,.6) -- (0,.1); }
\,\rvert$, $\langle\,
\tikz[baseline={([yshift=-.5*10pt*0.6]current bounding box.center)},scale=0.6,decoration={markings, mark=at position 0.7 with {\arrow[scale=1.25,>=stealth]{>}}}]{ 
	\draw[postaction=decorate] (.6,0) -- (.1,0);
	\draw[postaction=decorate] (0,.1) -- (0,.6); }
\,\rvert$, $\langle\,
\tikz[baseline={([yshift=-.5*10pt*0.6]current bounding box.center)},scale=0.6,decoration={markings, mark=at position 0.7 with {\arrow[scale=1.25,>=stealth]{>}}}]{ 
	\draw[postaction=decorate] (.6,0) -- (.1,0);
	\draw[postaction=decorate] (0,.6) -- (0,.1); }
\,\rvert$ for the `outgoing' lines at the vertex. In the diagrammatic notation in \eqref{eq:R-matrix} one should think of time running along the diagonal from bottom left to top right. $R(s)$ contains the vertex weights for the `black' vertices and $R(-s)$ for the `white' vertices.

A row of the lattice is described by the staggered (row-to-row) \emph{monodromy matrix}
\begin{equation}
\begin{aligned}
T(s) \coloneqq \ &
\tikz[baseline={([yshift=-.5*10pt*0.6+6pt]current bounding box.center)},
scale=0.6,font=\scriptsize]{
	\draw (0,1) -- (3.3,1) (4.1,1) -- (6,1);
	\draw (1,0) node[below]{$1$} -- (1,2);
	\draw (2.5,0) node[below]{$2$} -- (2.5,2);
	\draw (5,0) node[below]{$L$} -- (5,2);
	\draw[fill=white]  (1,1) node{$s$} circle (.35);
	\draw[fill=white]  (2.5,1) node{$-s$} circle (.35);
	\draw[fill=white]  (5,1) node{$\pm s$} circle (.35);
	\foreach \y in {-1,...,1} \draw (3.7 +.2*\y,1) node{$\cdot\mathstrut$};
} \\
= \ & R_L(\pm s) \cdots \, R_2(-s) \, R_1(s) \ ,
\end{aligned}
\end{equation}
where $R_j$ contains the weights for the $j$th vertex in that row. It is customary to write $B(s)$ for the $2^L\times 2^L$ matrix sitting in the upper right quadrant of $T(s)$. This matrix accounts for a row of the staggered six-vertex model with arrows on the horizontal external edges pointing outwards as for DWBCs:
\begin{equation}\label{eq:B's}
B(s) = \
\tikz[baseline={([yshift=-.5*10pt*0.6+6pt]current bounding box.center)},
scale=0.6,font=\scriptsize,decoration={markings, mark=at position 0.625 with {\arrow[scale=1.25,>=stealth]{>}}}]{
	\draw[postaction=decorate] (1,1) -- (0,1); \draw (1,1) -- (3.7,1) (4.5,1) -- (5.7,1); \draw[postaction=decorate] (5.7,1) -- (6.7,1);
	\draw (1.35,0) node[below]{$1$} -- (1.35,2);
	\draw (2.85,0) node[below]{$2$} -- (2.85,2);
	\draw (5.35,0) node[below]{$L$} -- (5.35,2);
	\draw[fill=white]  (1.35,1) node{$s$} circle (.35);
	\draw[fill=white]  (2.85,1) node{$-s$} circle (.35);
	\draw[fill=white]  (5.35,1) node{$\pm s$} circle (.35);
	\foreach \y in {-1,...,1} \draw (4.1 +.2*\y,1) node{$\cdot\mathstrut$};
} \ \, .
\end{equation}

The staggered domain-wall partition function can then be expressed as an entry of a `staggered' product of $L$ such matrices~\footnote{The partition function for an $L\times L$ lattice with $L$~even and PBCs, cf.~\cite{KLDB_16}, is obtained by defining the staggered transfer matrix $t(s)\coloneqq \text{tr}\, T(s) = A(s) + D(s)$, where $A(s)$ and $D(s)$ are similar to \eqref{eq:B's} but with both arrows pointing left and right, respectively. Then the partition function is $Z_L^\text{PBC}(s) = \text{tr}\{[t(-s) \,t(s)]^{L/2}\}$.}:
\begin{equation} \label{eq:Z(s)}
Z_L(s) = \langle \Down \Down \cdots \Down \rvert \, B(\pm s) \cdots B(-s) \, B(s) \, \lvert \Up \Up \cdots \Up \rangle \ .
\end{equation}
For example, if $L=1$ then $B(s)$ is $\begin{pmatrix} \color{gray!85}{0} & \color{gray!85}{0} \\ e^s \, c & \color{gray!85}{0} \end{pmatrix}$ and $Z_1(s) = e^s \, c$. The ordinary domain-wall partition function is recovered in this algebraic language as $Z_L = Z_L(0)$. We have evaluated~\eqref{eq:Z(s)} for general~$s$ up to $L=12$, accounting for little over $10^{16}$ configurations.

To conclude this section let us comment on whether quantum integrability may be used to get some concise expression for $Z(s)$ valid for all $L$. The answer appears to be negative; at least the Korepin--Izergin approach mentioned in Sec.~\ref{sec:DW_partition_function} does not simply extend to $s>0$. Indeed, one can still write down four recursion relations obeyed by the inhomogeneous extension of \eqref{eq:Z(s)}, namely for $u_1 = v_1 -\gamma$, $u_1 = v_L+\gamma$, $u_L = v_1+\gamma$ and $u_L = v_L-\gamma$. However, for $s\neq0$ the inhomogeneous partition function is not symmetric in the $u_i$, so one does not get further Korepin-like recursion relations and the conditions do not uniquely determine $Z(s)$ for general $L$. The failure of $Z(s)$ to be symmetric in the $u_i$ is of course closely related to the fact that the staggered $R$-matrices~\eqref{eq:R-matrix} do not obey a Yang--Baxter equation --- even writing down the latter is problematic since the triangle featuring in that relation is not bipartite. The latter also obstructs the computation of $P_0$ using the so-called F-basis~\cite{MS_96}.

\section{Simulations}\label{sec:sec3}

Recall that the six-vertex model is equivalent to a height model known as the (body-centred) solid-on-solid model~\cite{Bei_77}. In this picture fixed boundary conditions ensure that the height of a configuration is bounded from below and above. Going around the boundary in some direction the DWBCs correspond to the height increasing along two opposite ends, say from $0$ to $L$, and then decreasing from $L$ back to $0$ along the other two ends. There are unique configurations of minimal and maximal height: the former corresponds to a valley of height~$0$ running along one diagonal, and the latter to a ridge of height~$L$ along the other diagonal. (Note that these are the ground-state configurations of the two FE~phases. The AF ground state corresponds to a diamond-shaped plateau, of height as close as possible to $L/2$, surrounded by steep slopes to the pits and peaks at the corners.) The existence of configurations of minimal and maximal height allows one to use \emph{coupling from the past} (CFTP) algorithms~\cite{PW_96,PW_98}, which ensure that one draws configurations from the equilibrium distribution making it a perfect simulation. Although CFTP can in principle be `shelled' around a variety of updating schemes, in practice it is only used in combination with local updates due to the difficulties that arise when the same global update needs to be performed on both the lower and higher configuration. In this work we prefer speed over sample accuracy as this allows us to investigate much larger systems, thus improving the reliability of our subsequent analysis of the thermodynamic limit. Rather than CFTP we thus use the full lattice multi-cluster algorithm~\cite{WSK_90}, as in \cite{KLDB_16}, with a reported dynamic exponent $z=0.005\pm0.022$ for PBCs~\cite{BN_98}, so that the correlation time can be considered independent of system size in practice. The accuracy of our simulations is checked in Sec.~\ref{sec:sec4} against the theoretical expressions that were reviewed in Sec.~\ref{sec:sec2}.

Our results are procured from Monte Carlo simulations using the full lattice multi-cluster algorithm in combination with parallel tempering~\cite{Par_92}. We use the multi-histogram method~\cite{VC_72,FS_89} to interpolate observables, the energy and staggered polarization in particular, in a temperature range around the critical temperature. The \textit{F}-model is well suited for both parallel tempering and the multi-histogram method as the specific heat is analytically known and bounded, cf.~\eqref{eq:e_cv_definition} below, such that a set of temperatures can be constructed a priori at which the energy distributions of `adjacent' configurations overlap significantly. Given a configuration at inverse temperature~$\beta$, its neighbouring configurations are set at $\beta'=\beta\pm\beta/\sqrt{C_v}$. In each simulation the acceptance probability of swapping two configurations is never less than~$47\%$. After each update a measurement is taken, with a minimum of $10^6$ measurements per system size per temperature, at up to $30$ different temperatures per system size. At each measurement we determine the total energy and staggered polarization, calculated based on the description in the first paragraph of Sec.~\ref{sec:P_0}, of the system as well as the local vertex density at each vertex in the system. In principle one can estimate the thermal average of any time-independent (local) observable that can be defined for the system, such as arrow correlations, in a similar fashion. Note that all cluster updates that would change the arrows on the boundary are rejected to preserve the DWBCs.

\section{Results}\label{sec:sec4}

\subsection{Energy and specific heat}

Unlike the energy, the partition function itself can not be directly measured in Monte Carlo simulations. Exceptions are very small systems ($L\leq 6$) for which our simulations happen to sample all microstates so that we are able to reconstruct the full staggered partition function. The resulting expressions for $E(\beta)= \langle E(C)\rangle$ and $P_0(\beta)$ precisely match those obtained via the QISM as described in Sec.~\ref{sec:P_0}. In general just a part of the phase space is sampled so the partition function cannot be reconstructed as the total energy~$E(\beta)$ is not known for all temperatures. However, the multi-histogram method allows us to use simulations done at finitely many temperatures to determine the partition function, up to an overall factor, on some finite temperature range.

\begin{figure}[h!] \centering
	\includegraphics[width=1.0\linewidth,clip=]{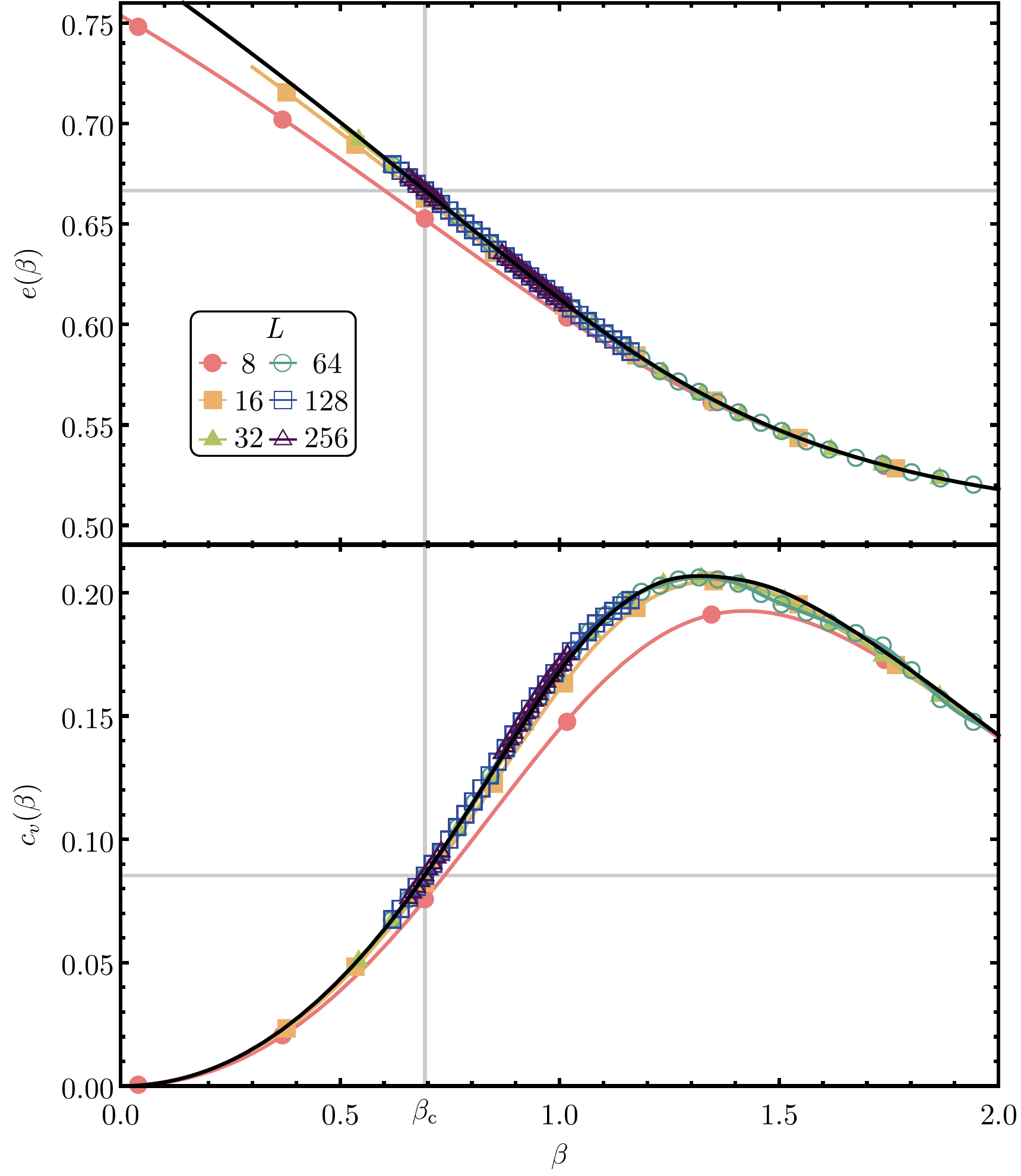}
	\caption{The average energy per site $e(\beta)$ (top) and specific heat per site $c_v(\beta)$ (bottom) as functions of inverse temperature~$\beta$. The critical points at $\beta_{\text{c}}$ are indicated by the gray lines. The black lines are the exact asymptotic expressions extracted from Eqs.~\eqref{eq:zD} and~\eqref{eq:zAF}. From the data points, indicating the temperatures at which the simulations are done, the coloured solid lines are calculated using the multi-histogram method. For both observables we see a convergence of the data to the analytically known expressions in the thermodynamic limit.}
    \label{fig:ECVcomparison}
\end{figure}

Fig.~\ref{fig:ECVcomparison} shows the energy per site $e(\beta)\coloneqq E(\beta)/L^2$ and the specific heat per site $c_v(\beta)\coloneqq C_v(\beta)/L^2$ as functions of inverse temperature. The simulation data are shown together with the exact expressions for infinite size extracted from Eqs.~\eqref{eq:zD} and~\eqref{eq:zAF} using
\begin{equation} \label{eq:e_cv_definition}
E(\beta) = -\frac{\partial \ln Z}{\partial \beta} \ , \quad
\frac{C_v(\beta)}{\beta^2} = -\frac{\partial E}{\partial \beta} = \frac{\partial^2 \ln Z}{\partial \beta^2} \ ,
\end{equation}
which yields $e(\beta_{\text{c}})=2/3$ and $c_v(\beta_{\text{c}})=8 \ln^2(2)/45$. We observe a convergence of the simulation data to the analytically known asymptotic values over all simulated temperature ranges.

\begin{figure}[h!]
	\centering
	\includegraphics[width=1.0\linewidth,clip=]{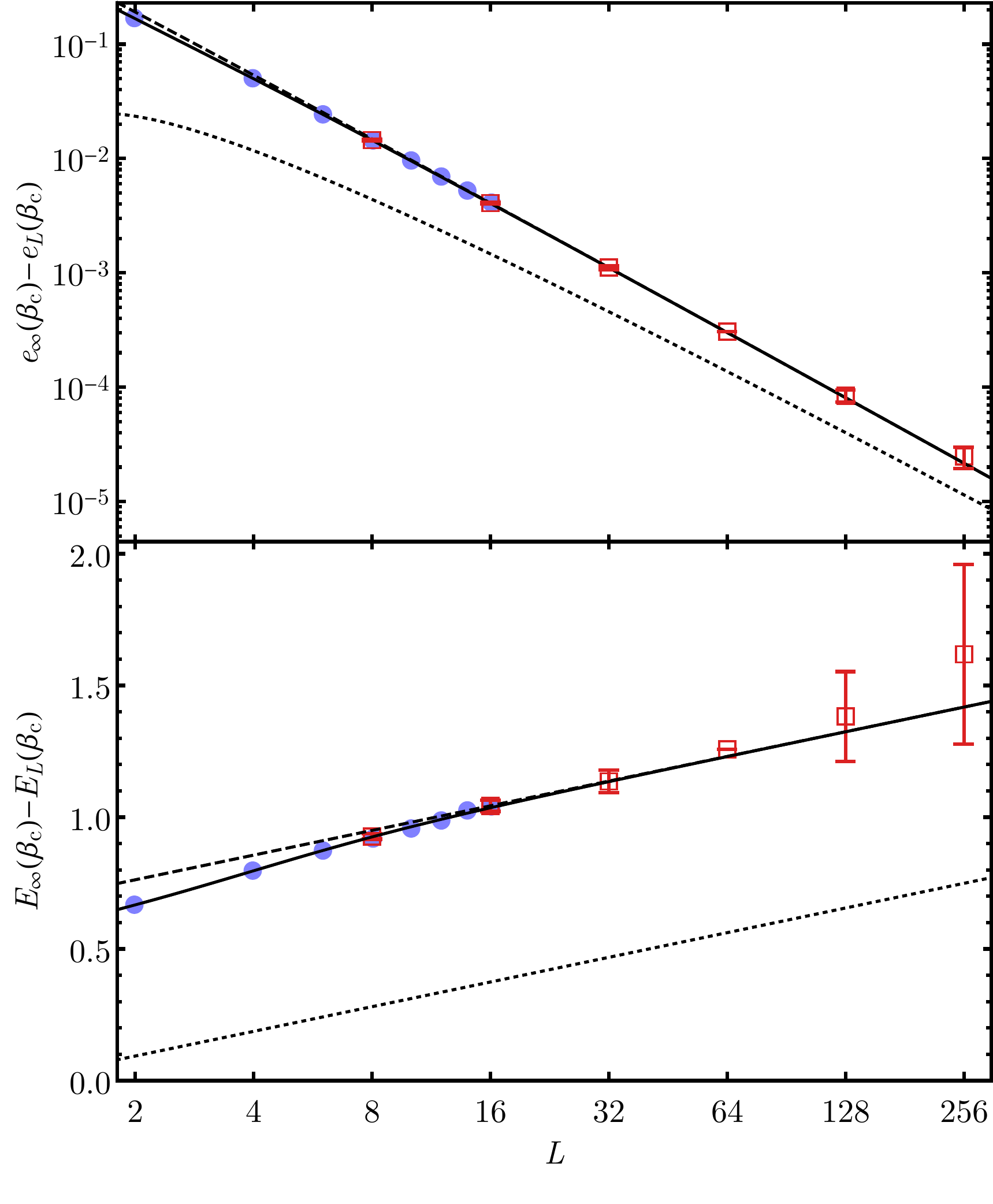}
	\caption{The difference between the energy per site (top) and total energy of the system (bottom), with $E_\infty(\beta_\text{c}) \coloneqq L^2 \, e_\infty(\beta_\text{c})=2\,L^2/3$, is shown as a function of system size. The solid blue disks represent the exact known values for small system sizes obtained from Eq.~\eqref{eq:Z_finite}. The open red squares denote best estimates obtained from our simulations. The error bars are estimates based on the fluctuations in the energy and the number of measurements taken. The expressions from Eq.~\eqref{eq:ecritpoint} with only the leading correction ($C_1\equiv C_2\equiv0$) and with first subleading correction ($C_1=0.669\pm0.018$), as well as the expression from Eq.~\eqref{eq:ecritpoint_best} ($C_2=1.6\pm1.2$, $C_3=14\pm12$, $\alpha=1.91\pm0.36$~\cite{Note7})
	are shown as dotted, dashed, and solid curves, respectively.}
	\label{fig:ElargeLbehaviour}
\end{figure}

To investigate the effects of the subleading corrections in the system size for the partition function further we focus on the critical point. Because of the smoothness of the partition function we can take the expressions for the disordered regime and evaluate them at the phase transition. Starting from Eq.~\eqref{eq:zD} we find the following expression for the energy per site $e_L(\beta_{\text{c}})$ at the critical point for system size~$L$:
\begin{align}\label{eq:ecritpoint}
e_L(\beta_{\text{c}}) = \frac{2}{3} - \frac{4}{3 \pi^2}\frac{\ln L}{L^2} - \frac{C_1}{L^2} + O(L^{-(\alpha+2)}) \ ,
\end{align}
with $C_1={-\!\lim}_{\gamma\to0^+} C_{\text{D}}^\prime(\gamma)/[\gamma\,C_{\text{D}}(\gamma)]$ an unknown parameter. Eq.~\eqref{eq:ecritpoint} can be checked against the expression for the energy derived directly from \eqref{eq:Z_finite} for small system sizes ($L\leq16$) as well as the simulation data for moderate system sizes. This is shown in Fig.~\ref{fig:ElargeLbehaviour} where $e_\infty(\beta_{\text{c}})-e_L(\beta_{\text{c}})$ and $E_\infty(\beta_{\text{c}})-E_L(\beta_{\text{c}})$ are plotted versus system size. The best unweighed fit, including only the asymptotically next-to-leading correction $C_1=0.669\pm0.019$, already shows very good agreement with both the exact and numerically obtained values. For $L\leq141$ this next-to-leading correction, $\sim 1/L^2$, is more important than the asymptotically leading correction, $\sim \ln L/L^2$.  This means that even at $L\sim10^{21}$ the two corrections in \eqref{eq:ecritpoint} just differ by a factor~$10$. Also note the high precision at which both the leading and first subleading corrections are measurable for systems as large as $L=256$, for which these corrections are of the order~$10^{-5}$.

A best estimate for the value of $\alpha$ can be found by assuming that the subleading terms in \eqref{eq:zD}, i.e.~the $O(L^{-\alpha})$, is of the form $g(\gamma)\, L^{-\alpha}$. This yields
\begin{align}\label{eq:ecritpoint_best}
e_L(\beta_{\text{c}}) \simeq \frac{2}{3} - \frac{4}{3 \pi^2}\frac{\ln L}{L^2} - \frac{C_1}{L^2} + \frac{C_2}{L^2\,(L^{\alpha}+C_3)} \ ,
\end{align}
where $C_2=\lim_{\gamma\to0^+}g^\prime(\gamma)/\gamma$ and $C_3=\lim_{\gamma\to0^+} g(\gamma)$ are again unknown. We assume that these limits make sense; the corrections are finite and must disappear for infinite systems. If we use our best value for $C_1$ and fit Eq.~\eqref{eq:ecritpoint_best} to the energies of small systems obtained from direct computation of~\eqref{eq:Z(s)}, see again Fig.~\ref{fig:ElargeLbehaviour}, the best estimates are $C_2=1.6\pm1.2$, $C_3=14\pm12$ and $\alpha=1.91\pm0.36$~\footnote{In fact, \cite[Thm.~6.1.3]{BL_13} shows that $\alpha\geq1$. Our value would mean that $b_1 = 0$ in the proof of that result. In principle it is possible to check this analytically, but the expression for $b_1$ is rather complicated. We thank P.~Bleher for correspondence about this point.}. The inclusion of these subleading correction does improve the fit qualitatively although error margins for best estimates of the parameters $C_2$ and $C_3$ are very large. With these values the crossover point where the terms proportional to $C_1$ and $C_2$ become comparable occurs already at $L=3.9$. The exact analytical values for the energy can be computed using \eqref{eq:Z_finite} or \eqref{eq:Z(s)}. We have done so for $L\leq16$; in either case most computation time was spent on the derivation of the energy from the partition function at the critical point rather than the calculation of the partition function itself.

\subsection{The logarithmic derivative of $P_0$}

\begin{figure}[b] \centering
	\includegraphics[width=1.0\linewidth,clip=]{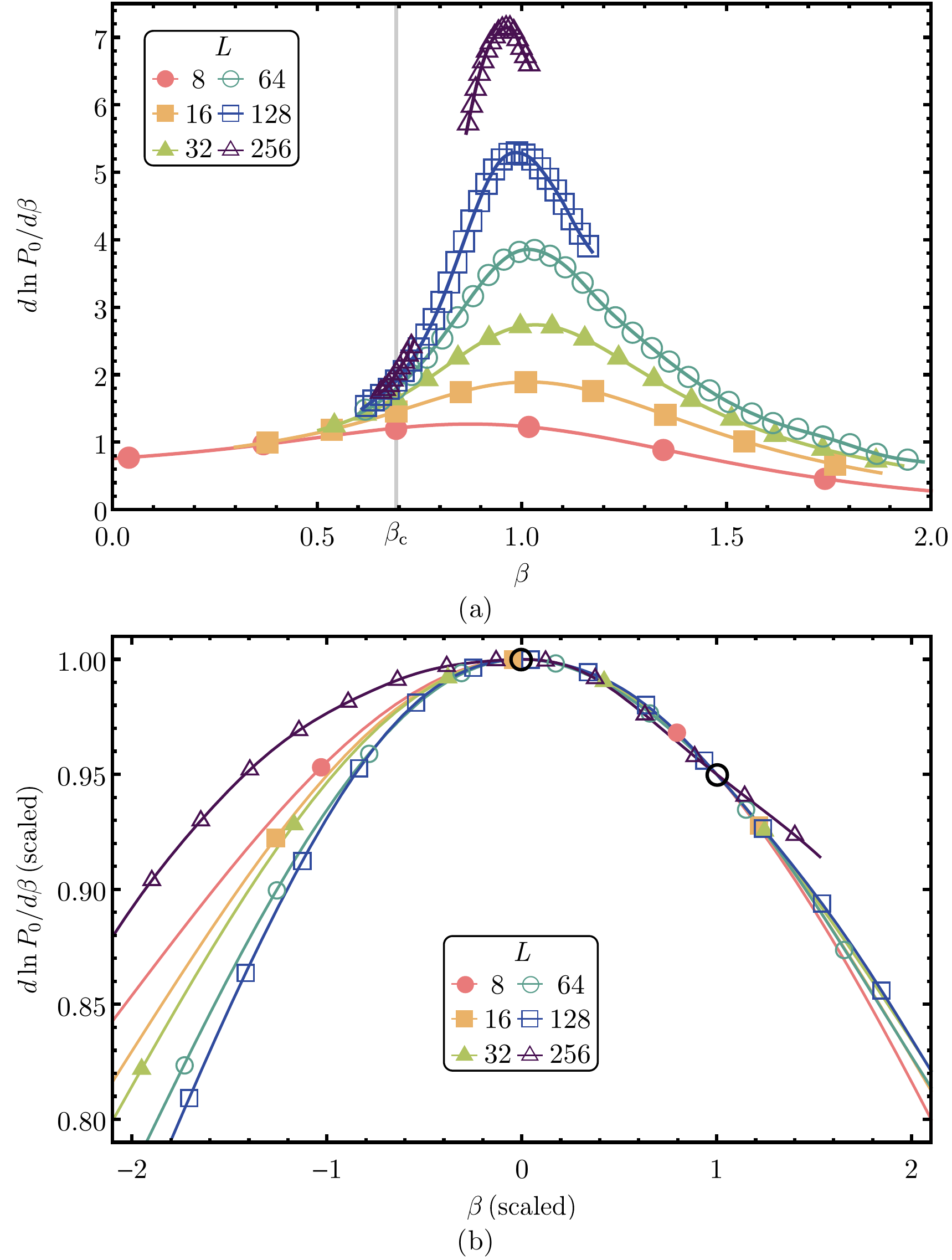}
	\caption{The observable $d \ln P_0 / d \beta$ is shown in (a) as a function of inverse temperature $\beta$ for various system sizes up to $L=256$. In the thermodynamic limit, under the assumption that $P_0$ is a valid order parameter, this function must have its peak at the critical point. In (b) the peak and the point where the curves attain $95\%$ of their peak height (at higher $\beta$) are scaled on top of each other, with the two points indicated by black circles. From this collapse we extract the peak position and typical width for further analysis, cf.~Fig.~\ref{fig:dlnP0dBETAextra}.}
    \label{fig:dlnP0dBETA}
\end{figure}

Similar to our work in~\cite{KLDB_16} we now study $d \ln P_0 / d \beta$, which must have a peak at the critical point for infinitely large systems if $P_0$ is a true observable of the infinite-order phase transition. As for the energy the multi-histogram method is used to obtain $d \ln P_0 / d \beta$ by interpolation between the temperatures at which the systems were simulated. Fig.~\ref{fig:dlnP0dBETA}~(a) shows $d \ln P_0 / d \beta$ as a function of inverse temperature for various system sizes up to linear size $L=256$. To obtain a numerical collapse for each system size we determine the peak coordinates $(\beta_{\text{max}},h_{\text{max}})$ as well as the typical width~$w$, which is defined as the absolute difference between $\beta_{\text{max}}$ and the lower temperature at which $d \ln P_0 / d \beta$ attains $95\%$ of the peak height. The numerical collapse is shown in Fig.~\ref{fig:dlnP0dBETA}~(b); unfortunately it is less clean than its counterpart for PBCs in~\cite{KLDB_16}.

\begin{figure}[ht!] \centering
	\includegraphics[width=1.0\linewidth,clip=]{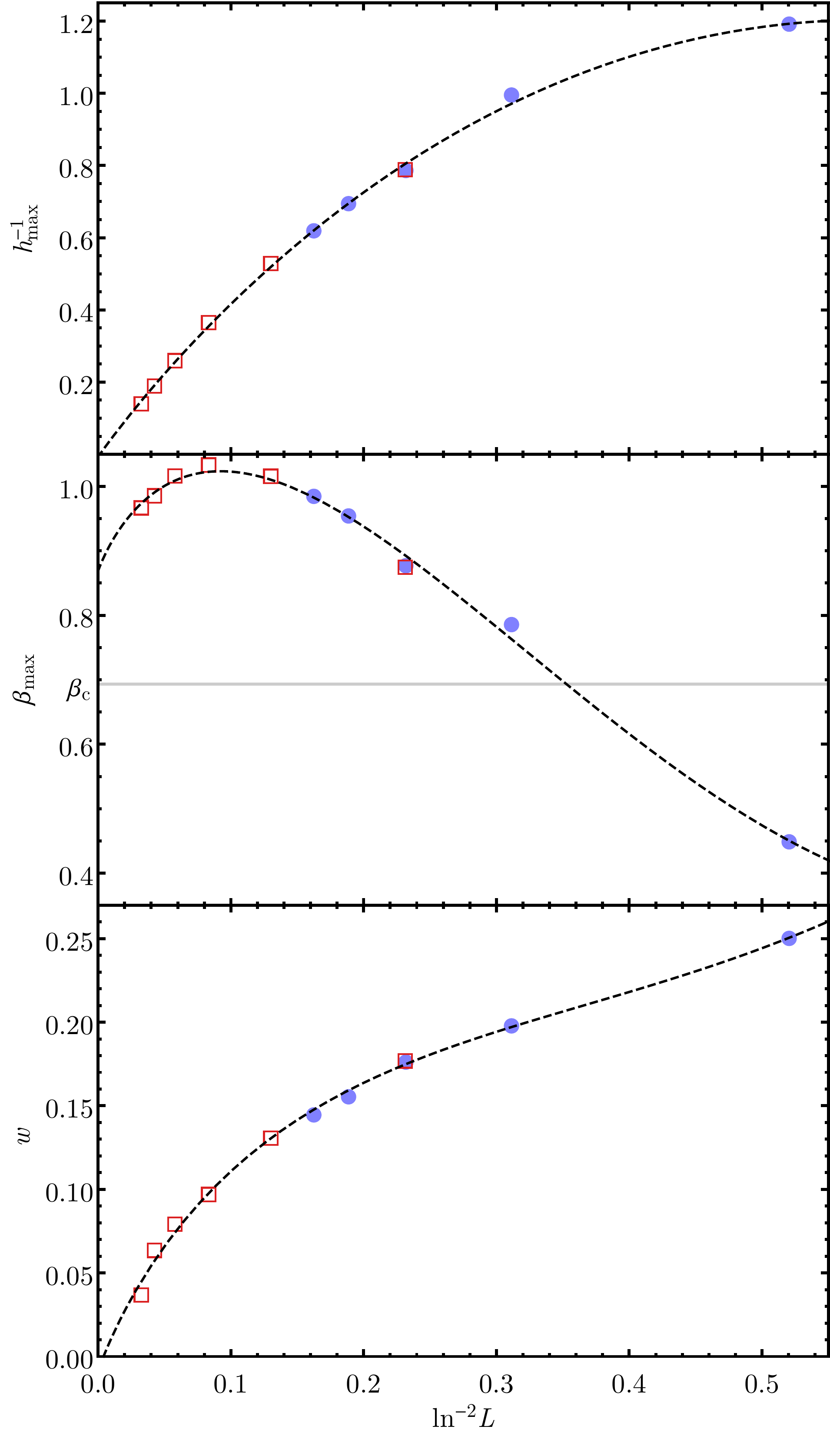}
	\caption{Values of the inverse peak height~$h_{\text{max}}^{-1}$ (upper panel), peak position~$\beta_{\text{max}}$ (central panel), and peak width~$w$ (lower panel) are shown for $d \ln P_0 / d \beta$ up to system size $L=256$ as a scaled function of $\ln^{-2} L$. Exact values for small system sizes obtained from Eq.~\eqref{eq:Z(s)} are shown as solid blue disks and best estimates obtained from our simulations as open red squares. The best unweighed fits of the form~\eqref{eq:dlnpdbscaling} are drawn as black dashed lines. For $\beta_{\text{max}}$ a fit through all data points results in a best estimate for the critical point $A_{\beta_{\text{max}}}=0.87\pm0.03$ far from the analytically known value~$\beta_{\text{c}}=\ln 2\approx 0.69$.
	For $h_{\text{max}}^{-1}$ ($A_{h_{\text{max}}^{-1}}=-0.01\pm0.03$, $B_{h_{\text{max}}^{-1}}=5.4\pm1.1$, $C_{h_{\text{max}}^{-1}}=-3\pm3$, $D_{h_{\text{max}}^{-1}}=-2\pm3$) and $w$ ($A_{w}=-0.009\pm0.009$, $B_{w}=2.4\pm0.4$, $C_{w}=-5\pm1$, $D_{w}=3.0\pm0.8$) the fits work well and are in agreement with $d \ln P_0 / d \beta$ becoming a Dirac delta-like distribution as $L\to\infty$.
	}
	\label{fig:dlnP0dBETAextra}
\end{figure}

Previously we found behavioural similarities between $d \ln P_0 / d \beta$ and the susceptibility~$\chi$ of the staggered polarization for PBCs~\cite{KLDB_16}. Since there are no known analytical expressions for the asymptotic behaviour of $P_0$ for DWBCs we fall back on the leading corrections known for PBCs~\cite{WJ_05}. In the case of PBCs the leading correction for the peak position of $\chi$ is of the form $\ln^{-2} L$, and so for DWBCs one could make the educated guess that the form of the peak of $d \ln P_0 / d \beta$ scales like
\begin{align}\label{eq:dlnpdbscaling}
x=A_{x}+B_{x}\ln^{-2}L+C_{x}\ln^{-3/2}L+D_{x}\ln^{-4}L ,
\end{align}
where $x$ is either the inverse peak height~$h_{\text{max}}^{-1}$, the peak width~$w$, or the position~$\beta_{\text{max}}$ of the peak. Fig.~\ref{fig:dlnP0dBETAextra} shows these quantities as a function of $\ln^{-2} L$ with the best fit of Eq.~\eqref{eq:dlnpdbscaling} to the three characteristics. The best estimates from an unweighed fit to all data points for the peak height are $A_{h_{\text{max}}^{-1}}=-0.01\pm0.03$, $B_{h_{\text{max}}^{-1}}=5.4\pm1.1$, $C_{h_{\text{max}}^{-1}}=-3\pm3$, and $D_{h_{\text{max}}^{-1}}=-2\pm3$. For the peak width the best estimates are given by $A_{w}=-0.009\pm0.009$, $B_{w}=2.4\pm0.4$, $C_{w}=-5\pm1$, and $D_{w}=3.0\pm0.8$. A similar fit for $\beta_{\text{max}}$ does not seem to work. Indeed, the best estimate for $A_{\beta_{\text{max}}}=0.83\pm0.02$ is far from the analytically known value $\beta_{\text{c}}=\ln 2$. Alternatively one could fix $A_{\beta_{\text{max}}}=\beta_{\text{c}}$, in which case the fit does not go through the data in a clean fashion. Although this method does not reliably give an estimate for the critical point it does show the convergence of $d \ln P_0 / d \beta$ to a Dirac delta-like distribution as the system size tends to infinity. From Fig.~\ref{fig:dlnP0dBETAextra} we see that in practice direct computation using~\eqref{eq:Z(s)} cannot be used outside of the regime in which subleading finite-size corrections are important. Simulations reveal the decrease in $\beta_{\text{max}}$ for increasing system size.

\subsection{Arctic curves}\label{sec:arctic}

So far we have investigated global quantities. For inhomogeneous (not translationally invariant) systems such as the \textit{F}-model with DWBCs such properties provide rather coarse information, as a lot of the local information is averaged away.

\begin{figure*} \centering
	\includegraphics[width=1.0\linewidth,clip=]{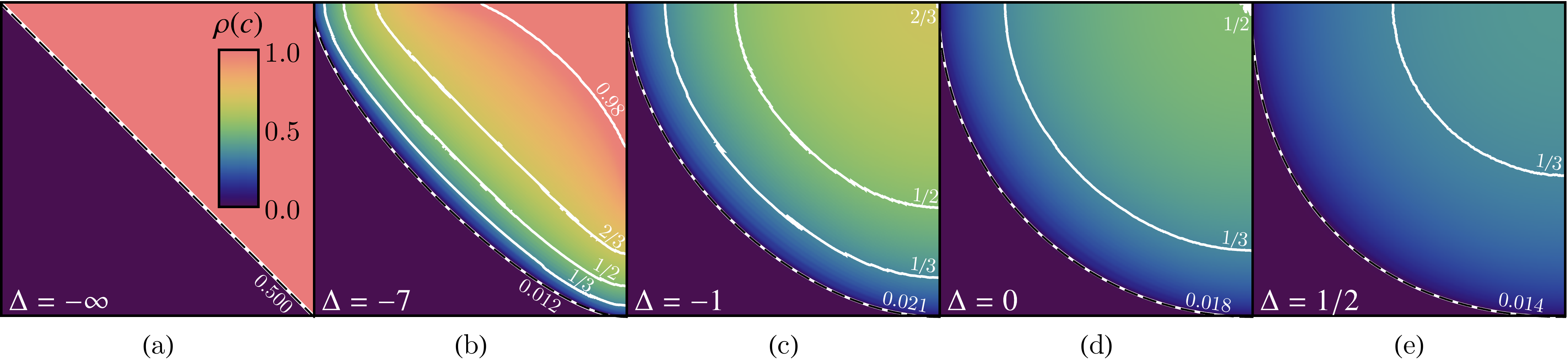}
	\caption{The thermally averaged density $\rho(c)$ of $c$-vertices at $L=512$ show phase separation at different temperatures. The density lies between zero (shown in purple) and one (light red). White solid contour lines are drawn at the values $1/3, 1/2, 2/3, \text{ and } 0.98$, as indicated, of $\rho(c)$. The outer white solid contours are drawn at temperature-dependent values of $\rho(c)$ (see Table~\ref{tab:table-values}) that give the best qualitative match with the arctic curves~\cite{Joh_05,FS_06,CP_10a,CP_10b,CPZ_10}, which are shown as dashed black curves. At zero temperature~(a) the AF region is a diamond. At slightly elevated temperatures~(b) the AF and FE regions are separated by a temperate region. As the temperature increases past its critical value~(c), at which the AF region disappears, the arctic curve deforms to a circle at the free-fermion point~(d). The system at infinite temperature is shown in~(e), in which the arctic curve is a sort of inflated circle, with the arcs deformed somewhat towards the corners of the domain.
	}
	\label{fig:Cdensity}
\end{figure*}

Fig.~\ref{fig:Cdensity} shows the thermally averaged $c$-vertex density~$\rho(c)$, together with several contour lines, for a system of linear size $L=512$ at various temperatures: zero temperature ($\beta\to\infty$, $\Delta\to{-\infty}$), below the critical point ($\beta=2 \beta_\text{c}$, $\Delta={-7}$), at the critical point ($\beta_\text{c}=\ln 2$, $\Delta={-1}$), at the free-fermion point ($\beta=\beta_\text{c}/2$, $\Delta=0$), and at infinite temperature ($\beta=0$, $\Delta=1/2$). For nonzero temperature 10 independent simulations, each yielding $10^6$ measurements, were performed per temperature to calculate the local vertex density. We use the global symmetries described in App.~\ref{sec:app} to get a smoother $\rho(c)$-profile by averaging at a given $\Delta$. At the centre $\rho(c)$ is always at a maximum. For zero temperature, the critical temperature, and the free-fermion point the maximal values are $1$ and about $2/3$ and $1/2$, respectively. At low temperatures there is a AF region, with constant $\rho(c)$ close to unity signalling its ordered nature. As the temperature rises from zero a temperate region emerges that encloses the central AF region, completely engulfing it at the critical point, cf.~\cite{AR_05}. The arctic curves, exactly known for $\Delta=-\infty$ and $\Delta=0$~\cite{Joh_05,FS_06} and conjectured for $\Delta<1$ \cite{CP_10a,CP_10b,CPZ_10} are also shown in Fig.~\ref{fig:Cdensity}. The outer contours are drawn at temperature-dependent values for $\rho(c)$ (see Table~\ref{tab:table-values}) chosen such that those contours are qualitatively comparable to the known and conjectured forms of the arctic curves. We see that our data match very well with the analytic expressions; for nonzero temperature the deviation from zero of the values given in Table~\ref{tab:table-values} is a measure of the influence of finite-size effects.

\begin{table}
	\begin{tabularx}{1.0\linewidth}{*6X} \toprule
		& (a) & (b) & (c) & (d) & (e) \\ \midrule
		$\beta$ & $\infty$ & $2\beta_\text{c}$ & $\beta_\text{c}{=}\ln2$ & $\beta_\text{c}/2$ & $0$ \\
		$\Delta$ & ${-\infty}$ & ${-7}$ & ${-1}$ & $0$ & $1/2$ \\ \midrule
		$\rho(c)$ & $0.500$ & $0.012$ & $0.021$ & $0.018$ & $0.014$ \\
		\bottomrule
		\hline
	\end{tabularx}
	\caption{The values for $\beta$ and $\Delta$ at which the simulations for Fig.~\ref{fig:Cdensity}~(a)--(e) were performed are given together with the values for $\rho(c)$ at which the outer contours are drawn. For finite $\Delta$ this gives a measure of the deviation from the asymptotic values $\rho(c)=0$ due to finite-size effects.}
	\label{tab:table-values}
\end{table}

\subsection{Oscillations in vertex densities}

\begin{figure}[h] \centering
	\includegraphics[width=1.0\linewidth,clip=]{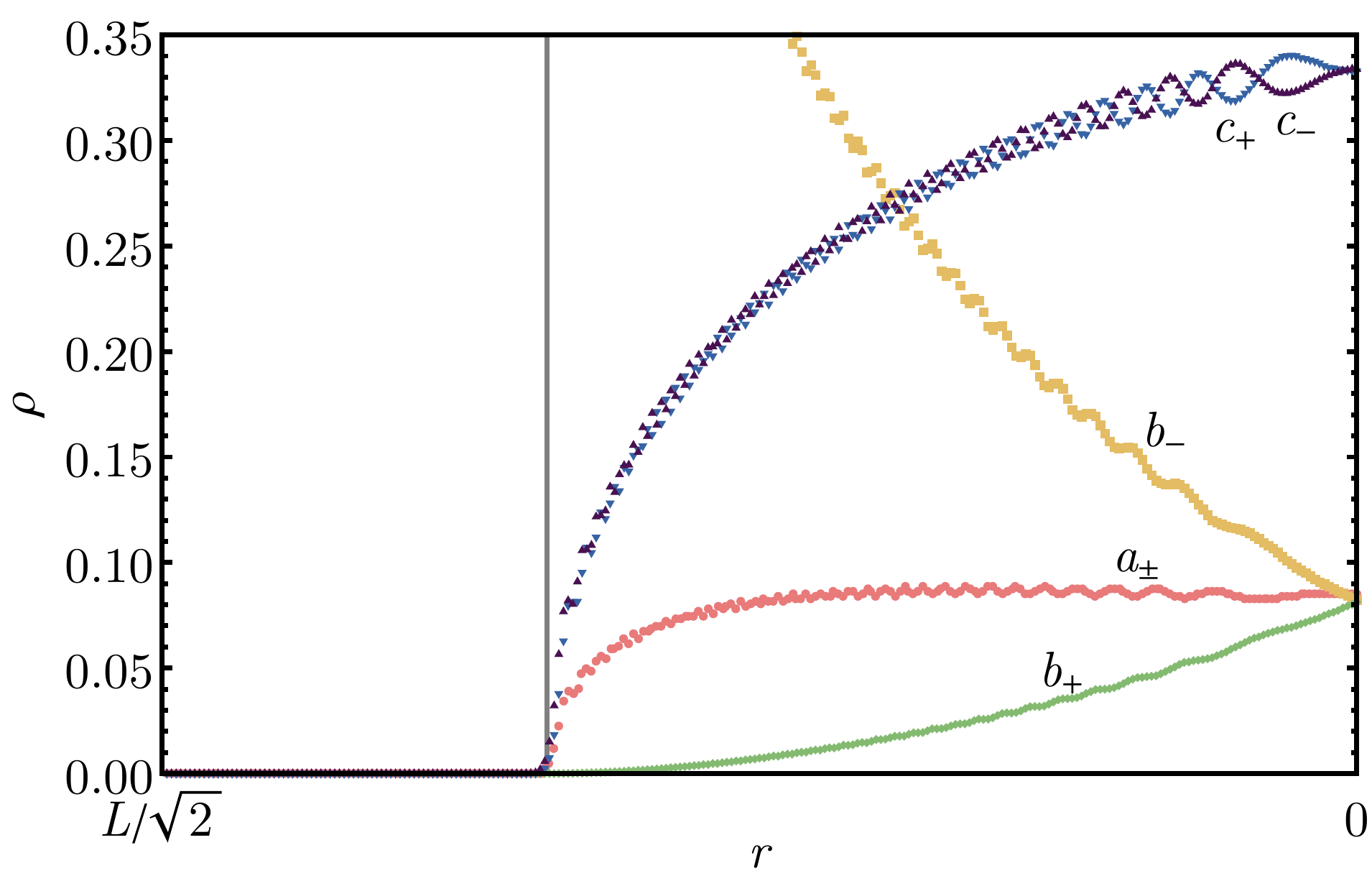}
	\caption{The thermally averaged densities~$\rho$ for all six vertices in a system of linear size $L=512$ at the critical point $\Delta=-1$ are shown along the diagonal from the $b_{-}$-dominated FE region to the centre at $r=0$. The grey vertical line marks the transition between the FE and temperate region~\cite{CPZ_10}. Each vertex density oscillates in the temperate region. Note that $\rho(c_\pm)$ are in antiphase around their average.}
	\label{fig:rhoallsection}
\end{figure}

Finally we turn to the structure inside the temperate region. In Fig.~\ref{fig:rhoallsection} we show the thermally averaged densities~$\rho$ along the diagonal from the FE region dominated by $b_{-}$-vertices ($r=L/\sqrt{2}$, bottom left corner in Fig.~\ref{fig:Cdensity}) to the centre ($r=0$) of a system of size $L=512$ at the critical point $\Delta=-1$. Along this diagonal one has $\rho(a_+)=\rho(a_-)$. Moreover if one considers $r$ to cover the full diagonal, ${-L}/\sqrt{2}\leq r \leq L/\sqrt{2}$, then $\rho(a_\pm)$ and $\rho(c_\pm)$ are even as functions of $r$ while $r\mapsto {-r}$ reverses $\rho(b_+)\leftrightarrow\rho(b_-)$. This once more allows us to exploit the global symmetries as explained in App.~\ref{sec:app} to average for the densities of $\rho(a_{\pm})$ and $\rho(b_{\pm})$ in Fig.~\ref{fig:rhoallsection}. Note that some of these transformations exchange $a_{\pm}\leftrightarrow b_{\pm}$ as they involve arrow reversal to preserve the boundary conditions. The supplementary material~\cite{Note2} shows the profiles of the six vertex densities for $L=100$ at different values of $\Delta$.

Using numerics, Sylju{\aa}sen and Zvonarev~\cite{SZ_04} first noticed oscillatory behaviour (`small wiggles') of the arrow polarization density for $\Delta<-1$, see Fig.~6 therein~\footnote{We thank I.~Lyberg for bringing this to our attention.}. Recently Lyberg \textit{et al}.~\cite{LKV_16u} recovered these oscillations while studying the local vertex densities exactly, cf.~the asymptotic expression of the arrow polarization found for $\Delta=0$ in~\cite{AD+_16}, as well as numerically. In Fig.~\ref{fig:rhoallsection} we observe oscillations for all of the vertex densities in the temperate region. The wavelengths of these oscillations are comparable functions of $r$ for each of the vertices. For lower temperatures these ripples are more pronounced yet the region in which they appear, viz.~the temperate region, becomes smaller. The thermally averaged densities~$\rho(c_+)$ and $\rho(c_-)$ are in antiphase (cf.~Fig.~\ref{fig:rhoallsection}) so these oscillations are masked if just $\rho(c)$ is considered as in Fig.~\ref{fig:Cdensity}. The complicated oscillatory behaviour in the temperate region can more clearly be seen from the thermally averaged $c$-vertex density difference $\delta\rho(c) \coloneqq \rho(c_-)-\rho(c_+)$. Let us emphasize that we focus on the density difference for the $c$-vertices because $\rho(c_\pm)$ are in anti-phase, so $\delta\rho(c)/2 = \rho(c)/2\, - \rho(c_+)$ allows us to study the oscillations of $\rho(c_+)$ about its `average' by approximating the latter with the average~$\rho(c)/2$ of $\rho(c_\pm)$. We should also point out that $\rho(c)$ itself exhibits oscillations, visible near the arctic curve for finite~$\Delta$ in Fig.~\ref{fig:Cdensity}; we have verified, however, that the $\rho(c)$- and $\delta\rho(c)$-oscillations have a phase difference of $\pi/2$, so the ripples in Fig.~\ref{fig:Cdensity} are related to the `FE oscillations' that we will introduce momentarily.

\begin{figure}[ht!] \centering
	\includegraphics[width=1.0\linewidth,clip=]{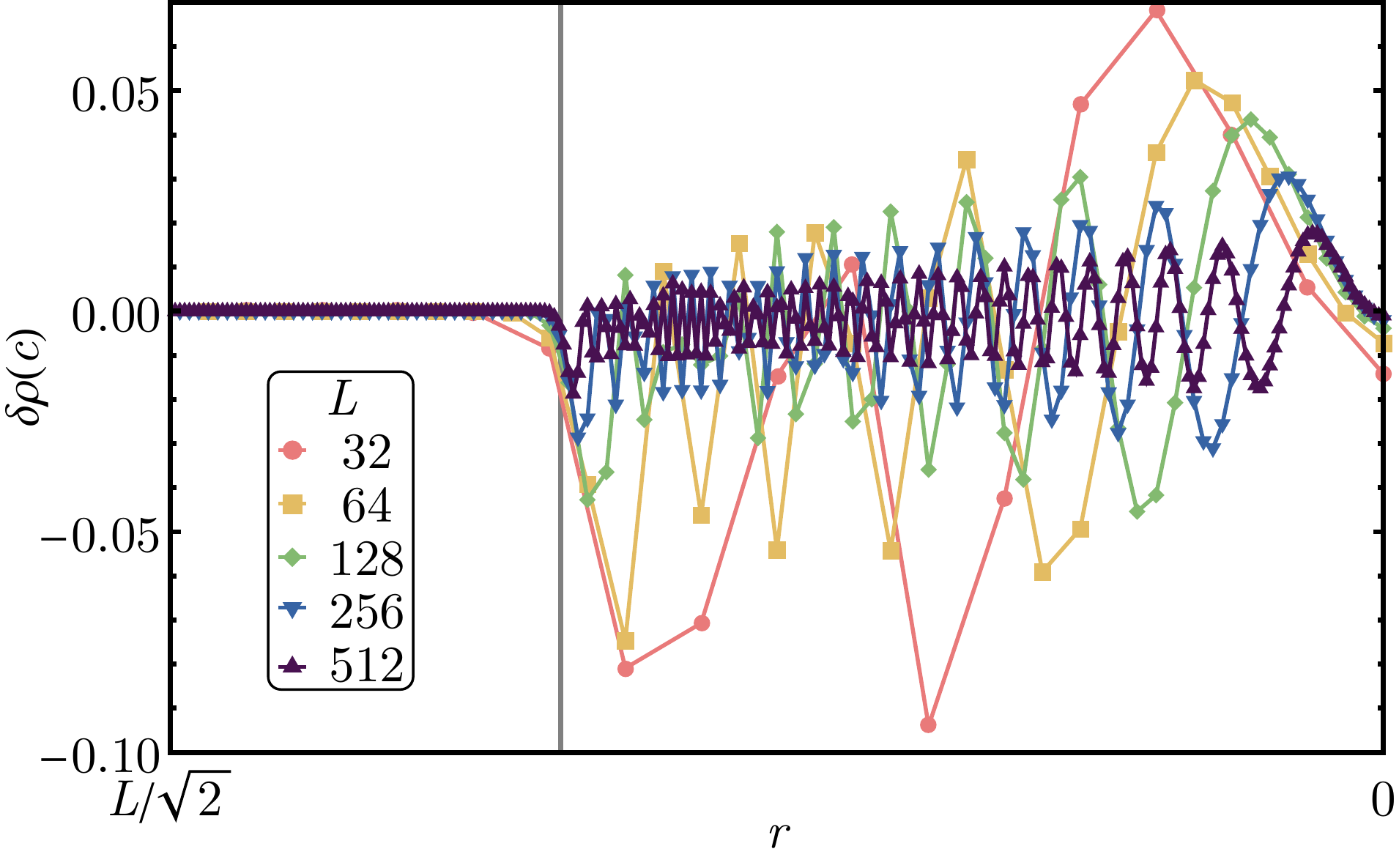}
	\caption{The difference $\delta\rho(c)$ is shown as a function of distance~$r$ along the diagonal to the centre (at $r=0$) for systems up to size~$L=512$ at~$\Delta={-1}$. The coloured lines are a guide to the eye and the grey vertical line denotes the transition between the FE-frozen and temperate region as in~\cite{CPZ_10}. The wavelength of the oscillations seems to increase sublinearly while the average wave amplitude decreases monotonically with system size, suggesting that this are finite-size effects.}
	\label{fig:deltaRHOscaling}
\end{figure}

To study the dependence on the system size of the oscillations in the temperate region Fig.~\ref{fig:deltaRHOscaling} shows $\delta\rho(c)$ along the diagonal for system sizes $L=32$ up to $L=512$. The wavelength of the oscillations is always largest at the edges of the temperate region. We observe a sublinear growth of the wavelength in~$L$. A best unweighed fit to the distance between the centre of the system ($r=0$) and the position of the maximum of~\ $\delta\rho(c)$ gives $(0.67\pm0.06) L^{(0.553\pm0.016)}$. Such a fit cannot be made for the maximal amplitude as our data are not accurate enough to distinguish between logarithmic or power-law behaviour. Still Fig.~\ref{fig:deltaRHOscaling} does clearly show that the average wave amplitude monotonically decreases with system size, suggesting that the oscillations are finite-size effects, as was conjectured in \cite{SZ_04}; cf.~Sec.~4 of \cite{LKV_16u}.

\begin{figure}[ht!] \centering
	\includegraphics[width=1.0\linewidth,clip=]{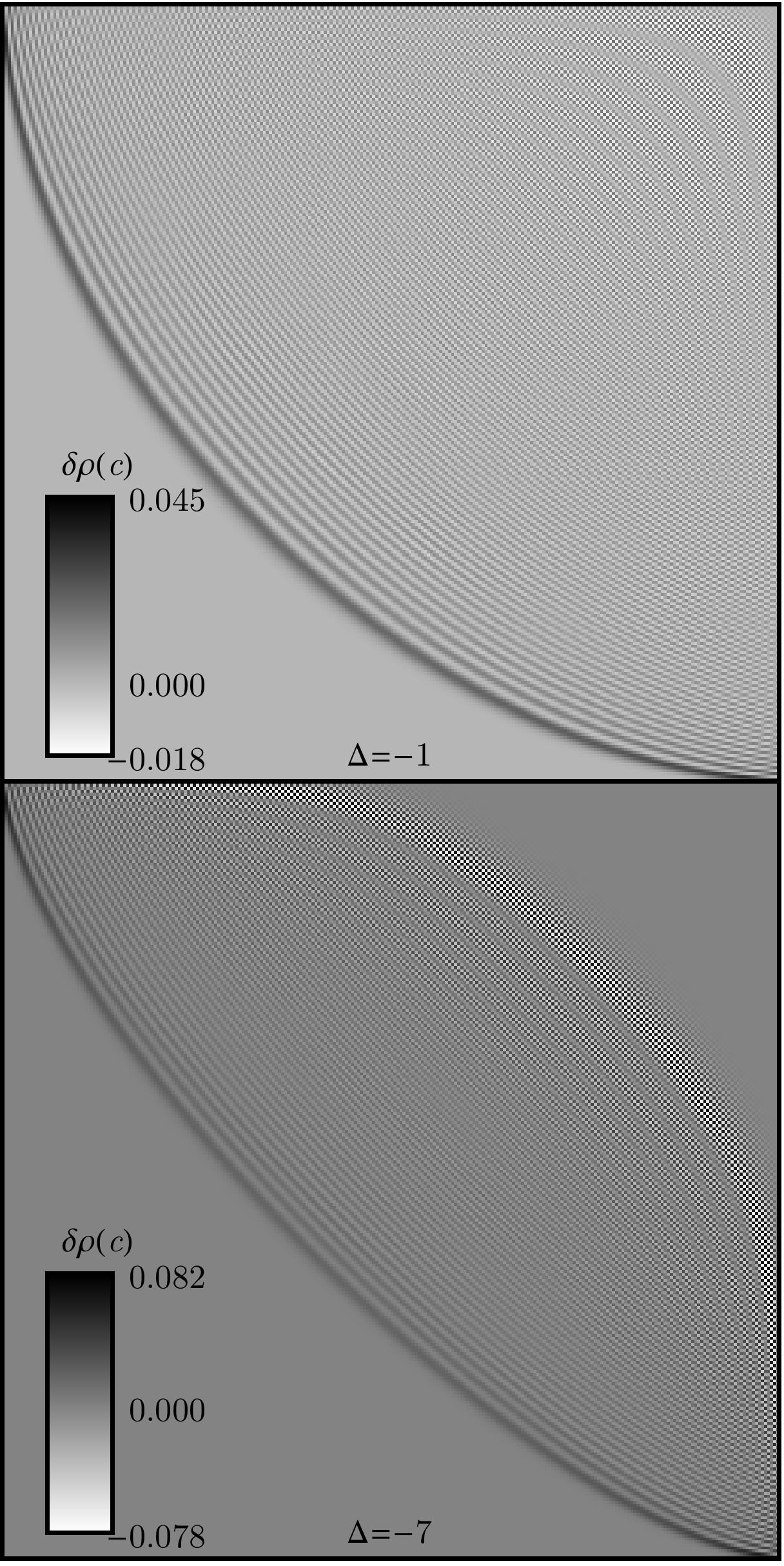}
	\caption{[High quality online.]
	The thermally averaged density difference $\delta\rho(c)$ is shown for a system of size $L=512$ at $\Delta=-1$ (upper panel) and $\Delta=-7$ (lower panel). Each pixel corresponds to a vertex. The FE-frozen region in the bottom left contains only $b_-$-vertices. Below the critical point the AF-frozen region appears (lower panel) in which $\delta\rho(c)=0$ due to the two-fold degeneracy for even $L$. Inside the intermediate temperate region at least two types of oscillations are visible. There appear to be chequerboard-like patterns in the `AF oscillations' even after thermal averaging, with opposite chequerboards in neighbouring oscillations. The `FE oscillations' are dominated by the vertices constituting the FE-frozen region (here $b_-$), with $\delta\rho(c)>0$ between them.}
	\label{fig:deltaRHOdensity}
\end{figure}

Fig.~\ref{fig:deltaRHOdensity} shows the profile of $\delta\rho(c)$ for systems at $\Delta={-1}$ and $\Delta={-7}$. Inside the temperate region there are at least two types of oscillations: one type, let us call them `AF oscillations', follows the boundary between the AF-frozen and temperate regions (which at $\Delta={-1}$ degenerates to the horizontal and vertical lines separating the quadrants), while the other type, `FE oscillations', follows the contours of the arctic curve between the temperate and FE-frozen regions. (Both of these types of oscillations may be discerned in \cite[Fig.~6]{LKV_16u} too, and the FE oscillations arguably already in Figs.~10 and~11 of \cite{SZ_04}. We should point out that in \cite{SZ_04} the term `AF oscillations' is instead used for the chequerboard patterns of $c_\pm$-vertices typical for AF order.)

Interestingly, upon closer inspection of Fig.~\ref{fig:deltaRHOdensity} we observe a chequerboard-like pattern inside the AF oscillations (cf.~\cite[Fig.~6, $\Delta={-10}$]{LKV_16u}), signalling site-to-site anti-correlations for $\rho(c_\pm)$ that persist over long distances along the oscillations, and justifying the name `AF' for these types of oscillations. Note that, albeit in a weaker form, these chequerboards survive thermal averaging: unlike the one in the AF region for even~$L$ it is a physical property of the system; see also the supplementary material to this work~\cite{Note2}. We observe that the chequerboards in adjacent oscillations are opposite, so the bands separating the oscillations can be understood as the result of destructive interference between the two chequerboards. Also note that such chequerboard-like anti-correlations are invisible when one focusses on the densities along the diagonal. Next we turn to the FE oscillations. The density profiles of all six vertices for $L=100$ can be found in the supplementary material~\cite{Note2}. The profile of $\rho(b_-)$ reveals that the interior of the FE oscillations near the frozen region dominated by $b_-$ are also dominated by $b_-$, and similar statements are true for the other quadrants. Fig.~\ref{fig:deltaRHOdensity} further shows that the regions between the FE oscillations are dominated by $c_-$-vertices ($\delta\rho(c)>0$) to account for the constraint $\#\,c_->\#\,c_+$ imposed by the DWBCs. Notice that as the FE oscillations approach the median, at the top of Fig.~\ref{fig:deltaRHOdensity}, they reduce to a chequerboard pattern on the median to merge with the interior of the largest AF oscillation.

To justify our observations let us explain in more detail how Fig.~\ref{fig:deltaRHOdensity} was obtained. We use the same data as for Fig.~\ref{fig:Cdensity}, based on 10 independent simulations each with $10^6$ measurements of local vertex density. 
We use the model's global symmetries to produce further configurations from those obtained from our simulations and sample over the full phase space as described in App.~\ref{sec:app}. Averaging over these configurations we obtain the profile for $\delta\rho(c)$ shown in Fig.~\ref{fig:deltaRHOdensity}, which correctly vanishes both in the FE and AF regions. For even as well as odd~$L$, however, the site-to-site anti-correlations between the AF oscillations in the \emph{temperate} region survive this averaging: unlike for the chequerboard in the AF region for even~$L$, this seems to be a \emph{statistical} property of the system. See also (the end of) App.~\ref{sec:app} and the supplementary material~\cite{Note2}.

\begin{figure}[ht!] \centering
	\includegraphics[width=1.0\linewidth,clip=]{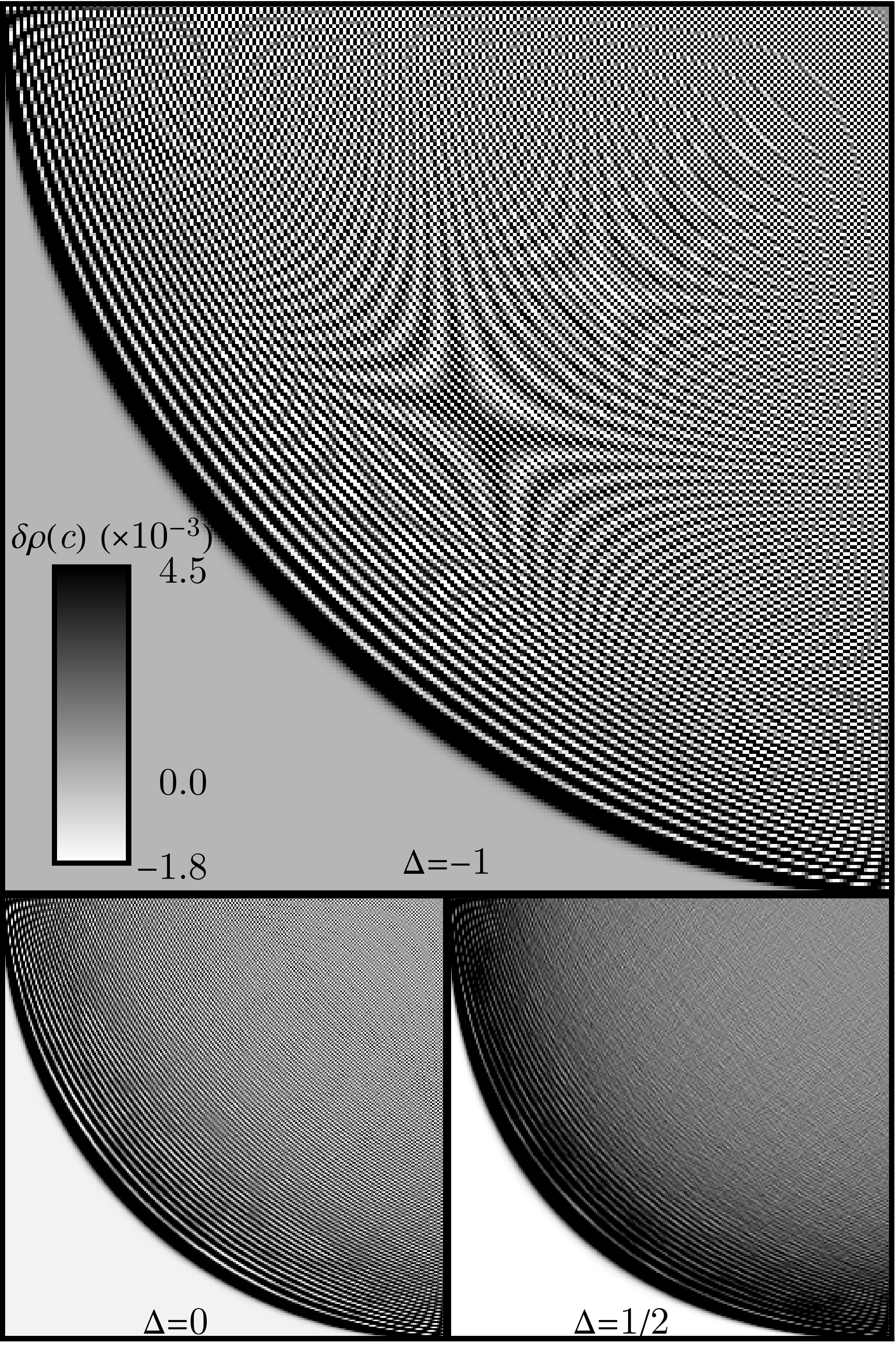}
	\caption{[High quality online.]
	The thermally averaged $c_\pm$-density difference $\delta\rho(c)$ for size $L=512$ at $\Delta=-1$ (upper panel), truncated at $10\%$ of the values from the upper panel in Fig.~\ref{fig:deltaRHOdensity}. Every pixel represents one vertex. This reveals weak `higher-order' oscillations in the temperate region with various saddle-point-like features; we can distinguish at least four of these along the diagonal, and more along the top and right. The lower panels show $\delta\rho(c)$ at $\Delta=0$ (left) and $\Delta=1/2$ (right), each again truncated at $10\%$ of its minimal and maximal value.}
	\label{fig:deltaRHOdensity_truncated}
\end{figure}

Besides the AF and FE oscillations following the curves separating the temperate and corresponding frozen regions there are also additional `higher-order' oscillations in $\delta\rho(c)$ that form intricate patterns in the temperate region that are barely visible in Fig.~\ref{fig:deltaRHOdensity}. To visualize these oscillations more clearly we truncate $\delta\rho(c)$ in the upper panel of Fig.~\ref{fig:deltaRHOdensity_truncated} at $10\%$ of the minimal and maximal values from the upper panel of Fig.~\ref{fig:deltaRHOdensity}. Even though the relative errors sometimes exceed the average value for very small $\delta \rho(c)$ the patterns exhibit a lot of structure, and cannot be attributed to random noise. These higher-order oscillations exhibit several saddle-point-like patterns around the centre of the temperate region. The structure is similar for lower $\Delta$; we have chosen $\Delta={-1}$ to get the largest temperate region. Some higher-order oscillations can be found in Fig.~7 of \cite{LKV_16u} for $\Delta=-10$.

The oscillations persist above the critical point. At $\Delta={-1/2}$ one can see FE oscillations in Fig.~6 of \cite{LKV_16u}. Going deeper into the D~phase the profiles of $\delta\rho(c)$ on the free-fermion line ($\Delta = 0$) and at the ice point ($\Delta = 1/2$) are shown in the lower panels of Fig.~\ref{fig:deltaRHOdensity_truncated}. At $\Delta = 0$ the FE oscillations are still clearly visible. Interestingly, even though the AF region has disappeared it leaves behind a `ghost' in the form of AF oscillations. Close inspection suggests there are higher-order oscillations too, with at least one saddle-point-like feature. At $\Delta = 1/2$ most structure of the temperate region is beyond the resolution of our data, yet one can still see weak FE oscillations as well as the tails of AF oscillations in the top-left and bottom-right corners of that panel.

\section{Summary and outlook}\label{sec:sec5}

In this work we have used Monte Carlo simulations to study the \textit{F}-model with DWBCs. Although a closed form for the partition function is analytically known for all system sizes in practice it is particularly useful for the exact computation of certain observables for fairly small systems and to obtain the asymptotic form and its finite-size corrections. Simulations allow for the investigation of systems of moderate size to complement such analytic results as well as to study properties that are not (yet) understood from an analytic point of view.

We have given best estimates for the parameters in the first three subleading finite-size corrections to the energy derived from the asymptotic partition function in Eq.~\eqref{eq:zD} at the critical point by fits to the average energies obtained from simulations. This tests the reliability of our simulations; they are precise enough to distinguish the different subleading corrections (Fig.~\ref{fig:ElargeLbehaviour}). The best estimates for the parameters suggest that the first subleading correction is non-negligible in comparison to the leading correction even for macroscopically sized systems, with $L\sim 10^{21}$. We find $\alpha=1.91\pm0.39$ for a previously unknown~\cite{Note7}
parameter in the asymptotic expression~\eqref{eq:zD} of the domain-wall partition function in the disordered regime found by Bleher and Fokin~\cite{BF_06}.

Following joint work with Duine and Barkema~\cite{KLDB_16} we have further investigated the order parameter based on the staggered polarization~$P_0$, of which we gave a description in the framework of the quantum-inverse scattering method (QISM). From a theoretical point of view it would be interesting to explore whether it is possible to adapt Baxter's work~\cite{Bax_73b} to obtain an exact expression for $P_0$ in the case of domain walls, at least in the thermodynamic limit, but we have not done so in the present work. If $P_0$ is a true order parameter for the model's IOPT, i.e.~it is constant on one side of the critical temperature and smoothly starts to change at the phase transition, then the observable $d \ln P_0 / d \beta$ must by definition have a divergence at the critical point for infinitely large systems. Using finite-size scaling, and extrapolating to the asymptotic case we have found that $d \ln P_0 / d \beta$ does indeed converge to a delta-distribution, see Fig.~\ref{fig:dlnP0dBETAextra}, although it fails to give an accurate estimate for the (analytically known) temperature at which the phase transition occurs. Of course the DWBCs together with the ice rule make the system that we have investigated rather special; the observable proposed in \cite{KLDB_16} may still be useful for the investigation of other models exhibiting an IOPT. One could also try using the susceptibility of $P_0$ instead; most of its peaks lie outside our simulation range, though the peaks that are visible appear to have a comparable quality for finite-size scaling.

In addition to these global (spatially averaged) properties we have studied local properties of the system. The profiles of the $c$-vertex density~$\rho(c)$ obtained for systems of size $L=512$ at various temperatures with $\Delta\leq1/2$ are shown in Fig.~\ref{fig:Cdensity}. In the antiferroelectric~(AF) phase our simulations corroborate the coexistence of three spatially separated phases as found in~\cite{SZ_04,AR_05}, with a flat central region exhibiting frozen AF order surrounded by a disordered~(D) `temperate' region and ferroelectrically~(FE) ordered corners. Our data agree very well with the arctic curves conjectured by Colomo and Pronko~\cite{CP_10b} and Colomo, Pronko and Zinn-Justin~\cite{CPZ_10}. It would be desirable to have similar analytic expressions for the `antartic curve' separating the temperate and AF-frozen regions for $\Delta<-1$.

Regarding the structure inside the temperate region our simulations confirm the oscillations first found by Sylju{\aa}sen and Zvonarev~\cite{SZ_04} and recently recovered by Lyberg \textit{et al}.~\cite{LKV_16u}. Our findings agree with those works, reproducing the patterns visible there, and uncover interesting additional features. Each vertex density oscillates with the same dependence of the wavelength on the position along the diagonal (Fig.~\ref{fig:rhoallsection}). Our data confirm the conjecture of \cite{SZ_04}, in accordance with \cite{LKV_16u}, that these oscillations are finite-size effects: their wavelengths appear to grow sublinearly --- roughly as $(0.67\pm0.06) L^{(0.553\pm0.016)}$ --- and their average amplitudes decrease with system size (Fig.~\ref{fig:deltaRHOscaling}). Our most detailed result regarding the structure of the temperate region are Figs.~\ref{fig:deltaRHOdensity} and~\ref{fig:deltaRHOdensity_truncated}. Here we have chosen to focus on the density difference for the $c$-vertices since $\rho(c_\pm)$ are in anti-phase (cf.~Fig.~\ref{fig:rhoallsection}), so $\delta\rho(c) \coloneqq \rho(c_-)-\rho(c_+)$ allows us to study the deviation of one type of vertex around its `average' without having to know an expression for the latter. We find several types of oscillations. The `AF' oscillations close to the AF-frozen region appear to be made up of chequerboards of $c_\pm$-vertices that (unlike the AF region in case of even~$L$) survive thermal averaging for even as well as odd~$L$, and are opposite between neighbouring oscillations. The `FE' oscillations near the FE-frozen region are dominated by the vertices constituting that frozen region; in between these oscillations there is a surplus of the type of $c$-vertices favoured by the DWBCs. In addition there appear to be weak `higher-order' oscillations in $c_\pm$-densities, forming various saddle-point-like patterns. The oscillations seem to grow weaker as $\Delta$ increases. Nevertheless the oscillations persist well into the D~phase, with FE and AF oscillations remaining partially visible at $\Delta=1/2$ (Fig.~\ref{fig:deltaRHOdensity_truncated}). A more quantitative understanding of these vertex-density oscillations and arrow correlations in the temperate region is desirable, both via simulations and through the analytic methods of \cite{Joh_05,FS_06}, \cite{CP_15} or \cite{AD+_16}. In fact, similar finite-size oscillatory behaviour is known to occur for the eigenvalue distributions in random-matrix models~\footnote{We thank K.~Johansson for pointing this out to us.}, see e.g.~\cite{EL_15}; this might shed light on the oscillations at least for $\Delta=0$, cf.~\cite{Joh_05,FS_06}.

In the near future we plan to report on phase coexistence, arctic-curve phenomena and the structure of the D~region for various other choices of boundary conditions, cf.~\cite{Kup_02}. Another interesting direction is the study the case of quantum-integrable `solid-on-solid' (SOS) models, with weights associated to the dynamical Yang--Baxter equation. The trigonometric SOS model is a one-parameter extension of the six-vertex model, and it would be interesting to understand the dependence of those phenomena on the additional `dynamical' or `height' parameter. It would also be very exciting if the theoretical and numerical investigations of the \textit{F}-model with domain walls would be complemented by experimental work as in e.g.~\cite{NMS_13}.

\section{Acknowledgements}\label{sec:sec6}

We thank P.~Zinn-Justin for bringing CFTP under our attention, G.~Barkema and F.~Colomo for feedback on earlier versions of this work, P.~Bleher for correspondence, and C.~Hagendorf, I.~Lyberg, and K.~Johansson for useful discussions. We also thank the two anonymous referees for their feedback. We are grateful to the Institute for Theoretical Physics at Utrecht University for the hospitality during the course of this work. JL gratefully acknowledges support from the Knut and Alice Wallenberg Foundation (KAW). This work is part of the D-ITP consortium, a program of the Netherlands Organisation for Scientific Research (NWO) that is funded by the Dutch Ministry of Education, Culture and Science (OCW).

\appendix

\section{Relating configurations with \\ opposite chequerboards in the AF region} \label{sec:app}

In this appendix we show that the \textit{F}-model has symmetries that can be used to sample the whole of phase space starting from any initial configuration obeying the ice rule and DWBCs. We should emphasize that the symmetries we have in mind are symmetries of the model, not of the individual configurations.

We start locally, with the symmetries of the \textit{F}-model at the level of individual vertices shown in Fig.~\ref{fig:sixvertices}. Such local symmetries must certainly preserve the lattice near the vertex, i.e.~the vertex with its four surrounding edges, so we are led to the dihedral group~$D_4$ of symmetries of the square. Concretely it contains rotations over multiples of $\pi/2$ as well as reflections in the horizontal, vertical and (anti)diagonal line through the vertex. These operations clearly preserve the ice rule. In fact, when the edges carry arrows there is one more thing we can do that is compatible with the ice rule: reversing all arrows, yielding an action of $\mathbb{Z}_2$ that commutes with the $D_4$.

One can check the preceding operations change the vertex weights as follows:
\begin{align*}
	\text{reflect } \mspace{5mu}
		\tikz[baseline={([yshift=-.5*10pt*.5]current bounding box.center)}]{\draw[<->] (0,0) -- (0,.3535);}
	\mspace{5mu} \,: \quad & a_\pm \leftrightarrow b_\pm \, , \\
	\text{reflect } 
		\tikz[baseline={([yshift=-.5*10pt*.5]current bounding box.center)}]{\draw[<->] (0,0) -- (.3535,0);}
	\,: \quad & a_\pm \leftrightarrow b_\mp \, , \\
	\text{reflect } 
		\tikz[baseline={([yshift=-.5*10pt*.5]current bounding box.center)}]{\draw[<->] (0,0) -- (.25,.25);}
	\,: \quad & a_+ \leftrightarrow a_- \, , \quad c_+ \leftrightarrow c_- \, , \\
	\text{reflect }
		\tikz[baseline={([yshift=-.5*10pt*.5]current bounding box.center)}]{\draw[<->] (.25,0) -- (0,.25);}
	\,: \quad & b_+ \leftrightarrow b_- \, , \quad\, c_+ \leftrightarrow c_- \, , \\
	\text{rotate }
		\tikz[baseline={([yshift=-.5*10pt*.5]current bounding box.center)}]{\draw[densely dotted] (0,0) -- (.3535,0); \draw[->] (.2,0) to [bend right=45] (0,.2); \draw (0,0) -- (0,.3535); }
	\,: \quad & a_\pm \mapsto b_\mp \, , \quad \, b_\pm \mapsto a_\pm \, , \quad c_+ \leftrightarrow c_- \, , \\
	\text{reverse arrows}\,: \quad & a_+ \leftrightarrow a_- \, , \quad b_+ \leftrightarrow b_- \, , \quad c_+ \leftrightarrow c_- \, ,
\end{align*}
where for each reflection we omit the two weights it preserves. Notice that, when using arrows along the edges to represent the microscopic degrees of freedom, the \textit{F}-model may be characterized as the special case of the six-vertex model for which the vertex weights are invariant under rotations over $\pi/2$, and that they are then further invariant under all of $D_4 \times \mathbb{Z}_2$.

At the global level $D_4 \times \mathbb{Z}_2$ acts on the configurations, where $D_4$ acts by symmetries of the $L\times L$ lattice if we would forget about the arrows. Not all of these global maps are allowed, though. Regarding the operations corresponding to $D_4$ the DWBCs are only preserved by a subgroup isomorphic to $\mathbb{Z}_2 \times \mathbb{Z}_2$ corresponding to rotation over $\pi$ and reflection in the horizontal and vertical symmetry axes of the lattice. However, that the remaining operations in $D_4$ also preserve the DWBCs if we combine them with arrow reversal \footnote{Thus the full global symmetry group of the \textit{F}-model with DWBCs is a subgroup of $D_4 \times \mathbb{Z}_2$ isomorphic to $D_4$. Recall that $D_4$ has a presentation in terms of two generators, $r$ and $s$, subject to the relations $r^4 = s^2 = (s\,r)^2 = e$. Concretely, $r$ acts by a rotation over $\pi/2$ while $s$ acts by a reflection. Write $g^*$ for the combination of $g\in D_4$ with arrow reversal. Then the subgroup of global symmetries is generated by $r^*$ and $s$, where the latter acts by reflection in the horizontal or vertical axes; clearly $(r^*)^4 = s^2 = (s\,r^*)^2 = e$. See also \eqref{model's_symmetries}.}.

The next question is how these operations act at the level of configurations. Recall that there are two AF ground states, with opposite chequerboard patterns for the alternating $c_+$- and $c_-$-vertices constituting the AF region; let us call them `$0$' and~`$1$'. Below the critical temperature ($\Delta<-1$) any configuration is closer (more similar) to one of these two ground states. Accordingly, the phase space decomposes into two parts, say $\mathcal{C}_i$, with $i\in\mathcal{C}_i$ for $i=0,1$. (See also the supplementary material.~\cite{Note2}) For sufficiently low temperatures (or $\Delta$) and large enough~$L$ it costs a macroscopically large amount of energy to go from the energetically favourable part of $\mathcal{C}_0$, i.e.~configurations close enough to~$0$, to the corresponding part of $\mathcal{C}_1$: the system is practically trapped in one of these parts. Since we start our Monte-Carlo algorithm from one of the two AF ground states we thus expect to stay in the corresponding part of the phase space as the system thermalizes for $\Delta <-1$ and large enough $L$.

Now we return to the model's symmetries. Consider the two AF ground states, $0$ and~$1$. When $L$ is even the four axes of reflection symmetry meet in the middle of the central face of the lattice, and it follows that the model's symmetries fall into two classes:
\begin{equation} \label{model's_symmetries}
	\begin{aligned}
	\text{fixing }i: \qquad & \text{identity} \, , \quad
		\tikz[baseline={([yshift=-.5*10pt*.5]current bounding box.center)}]{\draw[densely dotted] (0,0) -- (.3535,0); \draw[->] (.2,0) to [bend right=45] (0,.2) to [bend right=45] (-.2,0); \draw (0,0) -- (-.3535,0); }
	\ , \quad 
		\tikz[baseline={([yshift=-.5*10pt*.5]current bounding box.center)}]{\draw[<->] (0,0) -- (.25,.25);}
	\ {}^* \ , \quad
		\tikz[baseline={([yshift=-.5*10pt*.5]current bounding box.center)}]{\draw[<->] (.25,0) -- (0,.25);}
	{}^* \ , \\
	0\leftrightarrow 1\, : \qquad &
		\tikz[baseline={([yshift=-.5*10pt*.5]current bounding box.center)}]{\draw[densely dotted] (0,0) -- (.3535,0); \draw[->] (.2,0) to [bend right=45] (0,.2); \draw (0,0) -- (0,.3535); }
	{}^* \, , \quad
		\tikz[baseline={([yshift=-.5*10pt*.5]current bounding box.center)}]{\draw[densely dotted] (0,0) -- (.3535,0); \draw[->] (.2,0) to [bend left=45] (0,-.2); \draw (0,0) -- (0,-.3535); }
	\ {}^* \, , \quad 
		\tikz[baseline={([yshift=-.5*10pt*.8]current bounding box.center)}]{\draw[<->] (0,0) -- (.3535,0);}
	\, , \quad 
		\tikz[baseline={([yshift=-.5*10pt*.5]current bounding box.center)}]{\draw[<->] (0,0) -- (0,.3535);}
	\ ,
	\end{aligned}
\end{equation}
where `${}^*$' means combination with arrow reversal. More generally, \eqref{model's_symmetries} indicates how the model's global symmetries relate the $\mathcal{C}_i$.

Since for the \textit{F}-model these operations do not change the vertex weights, they preserve the energy of the configurations. Given any configuration we can act by the model's symmetries to generate further configurations of the same energy; we get up to eight configurations in this way, though it may be only four or two if the original configuration happened to possess some amount of symmetry. (One should really check for such symmetries of the original configuration to avoid overcounting, but at high enough $L$ we can skip this step as such symmetric configurations make up a negligible portion of the phase space.) Half of the configurations we get in this way lie in $\mathcal{C}_0$ and the other half in $\mathcal{C}_1$. The upshot is that after having run the Monte Carlo simulation we can use the model's symmetries to sample the full phase space, even from simulations that correctly sample around one of the two ground states.

\bibliography{references}

\end{document}